\documentclass[lettersize,journal,hidelinks]{IEEEtran}

\usepackage{cite}
\usepackage{algorithm}
\usepackage{algorithmic}
\usepackage{textcomp}
\usepackage{amsmath}
\usepackage{amssymb}
\usepackage{mathrsfs}
\usepackage{epsfig}
\usepackage{epsf}
\usepackage{theorem}
\usepackage[active]{srcltx}
\usepackage{enumerate}
\usepackage{latexsym,calc,dsfont,array,pgf,setspace}
\usepackage{acro}
\usepackage{tikz}
\usetikzlibrary{calc}
\usepackage{pifont}
\usepackage{subfigure}
\usepackage{epstopdf}
\usepackage{amsfonts}
\usepackage{float}
\usepackage{xcolor}
\usepackage{afterpage}
\usepackage[inkscapeformat=png]{svg}
\usepackage{graphicx}
\usepackage[acronym,shortcuts]{glossaries}
\usepackage{xcolor}
\usepackage{titlesec}
\usepackage{url}
\usepackage{orcidlink}

\newacronym{FH}{FH}{frequency-hopping}
\newacronym{JRC}{JRC}{joint radar and communication}
\newacronym{MIMO}{MIMO}{multiple-input and multiple-output}
\newacronym{SIMO}{SIMO}{single-input and multiple-output}
\newacronym{DFRC}{DFRC}{dual-function radar communications}
\newacronym{AF}{AF}{ambiguity function}
\newacronym{PRI}{PRI}{pulse repetition interval}
\newacronym{PRF}{PRF}{pulse repetition frequency}
\newacronym{PPH}{PPH}{polynomial-phase hopping}
\newacronym{QSM}{QSM}{quadrature spatial modulation}
\newacronym{ISAC}{ISAC}{integrated sensing and communications}
\newacronym{AWGN}{AWGN}{additive white Gaussian noise}
\newacronym{CPM}{CPM}{continuous phase modulation}
\newacronym{ECCM}{ECCM}{electronic counter-countermeasures}
\newacronym{RFA}{RFA}{random frequency agility}
\newacronym{RPA}{RPA}{random \ac{PRI} agility}
\newacronym{RFPA}{RFPA}{random frequency and \ac{PRI} agility}
\newacronym{QAM}{QAM}{quadrature amplitude modulation}
\newacronym{SINR}{SINR}{signal-to-interference pluse noise ratio}
\newacronym{SNR}{SNR}{signal-to-noise ratio}
\newacronym{ASK}{ASK}{amplitude shift keying}
\newacronym{PSK}{PSK}{phase shift keying}
\newacronym{ML}{ML}{maximum likelihood}
\newacronym{RIP}{RIP}{restricted isometry property}
\newacronym{OMP}{OMP}{orthogonal matching pursuit}
\newacronym{CIR}{CIR}{channel impulse response}
\newacronym{TDD}{TDD}{time division duplex}
\newacronym{CRKG}{CRKG}{channel reciprocity-based key generation}
\newacronym{FCM}{FCM}{Fuzzy C-means}
\newacronym{PIRS}{PIRS}{polar code-based information scheme}
\newacronym{BER}{BER}{bit error rate}
\newacronym{CRC}{CRC}{cyclic redundancy check}
\newacronym{BDR}{BDR}{bit disagreement rate}
\newacronym{SQ}{SQ}{scalar quantization}
\newacronym{VQ}{VQ}{vector quantization}
\newacronym{6G}{6G}{sixth-generation}
\newacronym{AI}{AI}{artificial intelligence}
\newacronym{V2V}{V2V}{vehicle to vehicle}
\newacronym{IM}{IM}{index modulation}
\newacronym{SM}{SM}{spatial modulation}
\newacronym{SIM}{SIM}{spatial index modulation}
\newacronym{PLS}{PLS}{physical layer security}
\newacronym{AoA}{AoA}{angle-of-arrival}
\newacronym{JCAS}{JCAS}{joint communication and sensing}
\newacronym{SotA}{SotA}{state-of-the-art}
\newacronym{FHCS}{FHCS}{frequency hopping code selection}
\newacronym{FHCSK}{FHCSK}{frequency hopping code-shift keying}
\newacronym{PH}{PH}{phase}
\newacronym{TX}{TX}{transmit}
\newacronym{AMP}{AMP}{amplitude}
\begin{document}

\title{Frequency Hopping Waveform Design for\\ Secure {Integrated Sensing and Communications}}

\author{
    Ali Khandan Boroujeni\textsuperscript{\orcidlink{0009-0003-5007-8535}},~\IEEEmembership{Graduate Student Member,~IEEE,} \\
    Giuseppe Thadeu Freitas de Abreu\textsuperscript{\orcidlink{0000-0002-5018-8174}},~\IEEEmembership{Senior Member, IEEE,}
    Stefan Köpsell\textsuperscript{\orcidlink{0000-0002-0466-562X}},~\IEEEmembership{Senior Member, IEEE,} \\
    Ghazal Bagheri\textsuperscript{\orcidlink{0009-0006-2740-8235}},~\IEEEmembership{Graduate Student Member,~IEEE,} \\
    Kuranage Roche Rayan Ranasinghe\textsuperscript{\orcidlink{0000-0002-6834-8877}},~\IEEEmembership{Graduate Student Member,~IEEE,} \\and
    Rafael F. Schaefer\textsuperscript{\orcidlink{0000-0002-1702-9075}},~\IEEEmembership{Senior Member, IEEE}
    \vspace{-2ex}
    \thanks{Ali Khandan Boroujeni, Stefan Köpsell, and Rafael F. Schaefer are with the Barkhausen Institut and Technische Universit\"at Dresden, 01067 Dresden, Germany (emails: ali.khandanboroujeni@barkhauseninstitut.org; \{stefan.koepsell,rafael.schaefer\}@tu-dresden.de).}
    \thanks{Giuseppe Thadeu Freitas de Abreu and Kuranage Roche Rayan Ranasinghe are with the School of Computer Science and Engineering, Constructor University (previously Jacobs University Bremen), Campus Ring 1, 28759 Bremen, Germany (emails: \{gabreu,kranasinghe\}@constructor.university).}
    \thanks{Ghazal Bagheri is with Technische Universit\"at Dresden, 01187 Dresden, Germany (email: ghazal.bagheri@tu-dresden.de).}
}

\markboth{}%
{Shell \MakeLowercase{\textit{et al.}}: Bare Demo of IEEEtran.cls for IEEE Journals}

\maketitle

\begin{abstract}
We introduce a comprehensive approach to enhance the security, privacy, and sensing capabilities of \ac{ISAC} systems by leveraging \ac{RFA} and \ac{RPA} techniques.
The combination of these techniques, which we refer to collectively as \ac{RFPA}, with \ac{CRKG} obfuscates both Doppler frequency and \acp{PRI}, significantly hindering {the chances that} passive adversaries {can successfully} estimate radar parameters.
In addition, a hybrid information embedding method integrating \ac{ASK}, \ac{PSK}, \ac{IM}, and \ac{SM} is incorporated { to increase the achievable bit rate of the system significantly. Next, a} sparse-matched filter receiver design is proposed to efficiently decode the embedded information with a low \ac{BER}.
Finally, a novel \ac{RFPA}-based secret generation scheme using \ac{CRKG} ensures secure code creation without a coordinating authority. 
The improved range and velocity estimation and reduced clutter effects achieved with the method {are} demonstrated via the evaluation of the \ac{AF} of the proposed waveforms.

\end{abstract}

% Note that keywords are not normally used for peer review papers.

\IEEEpeerreviewmaketitle

\glsresetall
%=====================
\vspace{-2ex}
\section{Introduction}
\IEEEPARstart{A} {\lowercase{great}} deal of effort has been made recently to develop \ac{ISAC} -- also referred to as \ac{JRC} -- systems \cite{Liu_JSC22, WangITJ2022, Wei_ITJ23,RanasingheTWC2024}, as the technology is recognized as one of the pillars of \ac{6G} wireless communications, expected to drive the creation of new markets \cite{ISACMarket2023} by enabling many new applications.
One aspect of the \ac{ISAC} paradigm which {despite} its importance has received comparatively less attention, however, is the implication that this technology might have on the privacy and security of users \cite{KaiqianISAC2023}.
Indeed, it is easy to foresee, especially when considering the concomitant development of \ac{AI}, how exposed users might be once everyday wireless devices acquire the capability of extracting (possibly autonomously) sensitive, contextual, and behavioral information about them \cite{LiComMag2023, ZhangNetwork2024, ChenWCom2023}.

{Given} such potential threats, \ac{ISAC} techniques incorporating security and privacy features have started to emerge \cite{SuTWC_2021, YuSecV2XISAC2023, GunluJSAIT2023}, giving rise to the notion of secure \ac{ISAC}.
Since security measures for the communication aspect of \ac{ISAC} {have already} been (and {continue} to be) thoroughly investigated \cite{ylianttila6GSec2020, MucchiOJCS2021, LiuITJ2021, KatsukiTIFS2023}, we hereafter focus on the secure mechanisms for the sensing part of \ac{ISAC}, ensuring it also meets communication security requirements.
{Among the v}arious approaches to integrate security and privacy {into} the sensing component of \ac{ISAC} {is, for instance, the method} in \cite{SuTWC_2021}, {where}  a wiretap channel model for a dual-functional radar-communication system was introduced, acknowledging the potential for targets to eavesdrop.
By utilizing artificial noise and constructive interference, the contribution endeavors to decrease the \ac{SINR} at {specific} target locations.
The approach thereby does not address, however, the existing vulnerability to threats regarding the privacy of target locations.

{In light of} the latter, a versatile and increasingly popular mechanism to add a layer of security and privacy to \ac{ISAC} systems is to employ the \ac{FH} framework \cite{WuTWC2022} in signal design\footnote{Enabling secure communication-centric \ac{ISAC}\cite{RanasingheICASSP2024,RanasingheWCNC2025}, which is a separate problem, will be addressed in our future work.} {to prevent signals transmitted by {an} \ac{ISAC} system being} exploited by other (possibly) malicious devices.
But since this approach implicates, from a wider perspective, {the} design of purpose-built waveforms for \ac{ISAC}, it requires that various performance metrics and system features such as {data rate}, sensing {accuracy and} computational/hardware complexity be taken into account.

To cite a few relevant contributions in this area, the work in \cite{HoangWCL_2022} seeks to increase the data rate of \ac{FH}-based \ac{ISAC} systems by modulating information in both the frequency and duration of sub-pulses.
In turn, the methods in \cite{BaxterTSP_2022} and \cite{HassanienRC_2017} aim to accommodate various signaling strategies, including hybrid modulation schemes combining \ac{PSK}, \ac{IM}, and code selection using \ac{FH} \ac{MIMO} waveforms for \ac{DFRC} systems, enhancing data rates but introducing challenges such as spectral leakage and range sidelobes.

Focusing on sensing accuracy, the concept of ambiguity function analysis was extended in \cite{ChenTSP_2008} from \ac{SIMO} to \ac{MIMO} radar systems and utilizes orthogonal waveforms to enhance spatial resolution and its impact on range and Doppler resolution.
Following that line of work, analytical expressions for \ac{PRI} agile waveforms and \ac{AF} metrics of \ac{RFPA} signals are given in \cite{LongTAES_2021}, along with insights into \ac{RFPA} waveform design, which reveal tendencies for improved sidelobe suppression and ambiguity attenuation.
Finally, \cite{AngelosanteTSP_2010} introduces a sparse linear regression approach for improved hop timing estimation in \ac{FH} signals, outperforming spectrogram-based methods, crucial for both \ac{FH} and \ac{PPH} signals.

{ 
Despite the progress made by works such as those aforementioned, several limitations remain which need be addressed. 
For instance, the artificial noise and constructive interference techniques used in \cite{SuTWC_2021} assume active attackers but overlook passive adversaries, which are more prevalent in practice.
In turn, the \ac{FH} framework in \cite{WuTWC2022} improves security but requires a challenging balance of performance metrics, complicating real-world implementation. 
And efforts to boost data rates, such as those in \cite{HoangWCL_2022, BaxterTSP_2022, HassanienRC_2017}, suffer from spectral leakage and sidelobe reduction, which degrade radar privacy and hinder target detection. 
In particular, \cite{ChenTSP_2008} and \cite{LongTAES_2021} reveal that optimizing sensing accuracy in \ac{MIMO} \ac{FH} systems remains computationally expensive, especially with multiple hopping frequencies, and involves trade-offs between range and Doppler resolution. 
Additionally, the \ac{IM} approach in \cite{BaxterTSP_2022} and \cite{HassanienRC_2017} raises data rates at the cost of an increase in the range sidelobes, compromising clutter suppression and target identification, not to mention that \ac{IM} symbol recovery typically has high computational costs, making real-time processing impractical. 
Finally, the hybrid approaches proposed in these works overlook the potential advantages of modulation schemes such as \ac{ASK}.
}

{
In response to the shortcomings identified above, this paper introduces a comprehensive and innovative framework for \ac{ISAC} systems. 
Our contributions provide a breakthrough in several key areas, including secure transmission, privacy amplification, hybrid secure \ac{TX} signal design and receiver development. 
These contributions also address significant challenges in modern \ac{ISAC} use cases, including communications and sensing security and privacy, spectral efficiency, \ac{BER}, and computational complexity. 
Below, we categorize our contributions into two primary areas.

% \vspace{{0.5em}}
\noindent\textbf{A. Novel Secure Hybrid \ac{ISAC} TX Signal Model}

In this area, we contribute a hybrid transmit signal model that addresses the \ac{ISAC} functionalities as follows:
\vspace{-0.5ex}
\begin{itemize}
\item \textbf{Modification of \ac{RFA} and \ac{RPA} for \ac{ISAC} Platforms:}
\ac{RFA} and \ac{RPA} are adapted for \ac{ISAC}, introducing a secure hybrid modulation scheme combining \ac{ASK}, \ac{PSK}, \ac{SM}, and \ac{IM} with enhanced \ac{RFPA}. This improves spectral efficiency, radar performance, target detection, and resilience against adversarial attacks while securing the transmitter and communication receiver.

\item \textbf{Sparse Low-Complexity Receiver Design for Hybrid Modulation:}
A sparse-matched filter receiver decodes hybrid signals efficiently, reducing computational complexity and improving \ac{BER} in \ac{ISAC} systems.
\end{itemize}

\noindent\textbf{B. New Machine Learning-Based Vector Quantization for Shared Secret Generation}

% \vspace{-0.5ex}
In this category, a novel \ac{ML} technique enhances shared secret generation and utilization for secure communication.

% \vspace{-1.5ex}
\begin{itemize}
\item \textbf{New \ac{FCM} Vector Quantization Based on Reciprocal \ac{CIR}:}
A novel equal-sized \ac{FCM} vector quantization approach uses reciprocal \ac{CIR} of \ac{MIMO}-\ac{FH} channels, maximizing entropy, adapting to real-world channel conditions, enhancing shared secret accuracy, and mitigating information leakage.

\item \textbf{Novel Cluster Labeling Method for Overcoming Communication Overhead:}
A new cluster labeling method eliminates the need to transmit cluster information, reduces communication overhead, and preserves privacy during the quantization process.

\item \textbf{Creative Utilization of Shared Secrets as Pseudo-Random Sequences for \ac{RFA} and \ac{RPA} Techniques:}
The shared secrets derived from the \ac{FCM} approach are utilized as pseudo-random sequences for \ac{RFA} and \ac{RPA} at the physical layer, effectively integrating key generation into the security framework. This improves security and privacy in adversarial environments by obfuscating both the Doppler frequency and \ac{PRI}s, thereby significantly complicating passive adversaries' ability to estimate the radar's and target's location and velocity.
\end{itemize}
}

The remainder of the paper is organized as follows: 
Section \ref{Preliminaries} covers preliminaries, including the system and signal model and the calculation of the \ac{FH} ambiguity function. Section \ref{Waveform-Design} reviews the state of the art and introduces a new secure \ac{RFPA}-\ac{FH}-\ac{ISAC} signal model and its ambiguity function calculation. Section \ref{sec:Information-Embedding-Schemes} discusses various information embedding schemes and their receivers. The new \ac{RFPA} secret generation scheme is detailed in Section \ref{RFPA_Secret_Generation}. Section \ref{Complexity} addresses the complexity of the proposed algorithms. Finally, section \ref{Results} evaluates the algorithms' performance using communication, radar, and security metrics and compares their performances.

\vspace{-2ex}
\begin{table}[H]
\centering
\caption{Notations and Symbols Used in the Study}
\label{tab:notations}
\scriptsize
\setlength{\tabcolsep}{4pt} % Adjust column separation
\renewcommand{\arraystretch}{0.9} % Adjust row separation
\begin{tabular}{ll}
\textbf{Notation} & \textbf{Explanation} \\
\hline
$T_p$                         & Radar Pulse Repetition Interval (PRI) \\
$K$                           & Number of available frequency hops \\
$Q$                           & Number of sub-pulses per radar pulse \\
$BW$                          & Radar transmit bandwidth \\
$\Delta_f$                    & Radar sub-pulse frequency interval \\
$\Delta_t$                    & Radar sub-pulse duration \\
$f_l$                         & Carrier frequency of $l^{th}$ pulse \\
$T_l$                         & Starting point of pulse in the $l^{th}$ PRI \\
$J_{ASK}$                     & Size of ASK constellation \\
$J_{PSK}$                     & Size of PSK constellation \\
$\angle$                      & Phase indicator operator\\
$(\cdot)_q$                   & Sub-pulse $q$ \\
$\odot$                       & Hadamard product \\
$\mathbf a \oplus \mathbf b$  & XOR operation performed on bit strings \\ & of $\mathbf{a}$ and $\mathbf{b}$\\
$(\cdot)^*$                   & Complex conjugate \\
$(\cdot)^T$                   & Transpose \\
$(\cdot)^H$                   & Transpose and conjugate transpose \\
$\mathbf a \cdot \mathbf b$   & Dot product of two vectors $\mathbf a$ and $\mathbf b$\\
$\lfloor \cdot \rfloor$       & Floor function \\
$\mathbf{I}_M$                & M × M identity matrix \\
$\mathbf{1}_M$                & {V}ector of size $M$ consisting of all ones\\ 
{$\mathbf{A}^{\dagger}$}        & {Pseudo-inverse of $\mathbf{A}$, defined as $(\mathbf{A}^\mathrm{H} \mathbf{A})^{-1}\mathbf{A}^\mathrm{H}$}\\ 
{$[\cdot]^+$}                  & {$\max(0, \cdot)$}\\
$\text{diag}\{\mathbf{u}\}$   & Diagonal matrix with the main diagonal \\ & comprised of $\mathbf{u}$ \\
\hline
\end{tabular}
\end{table}
\vspace{-2.5ex}
%----------------------

\section{Preliminaries}
\label{Preliminaries}
\subsection{Wiretap Channel for \ac{ISAC} Model}

\begin{figure}[H]
\centering
\includegraphics[width=0.45\textwidth, height=0.23\textheight]{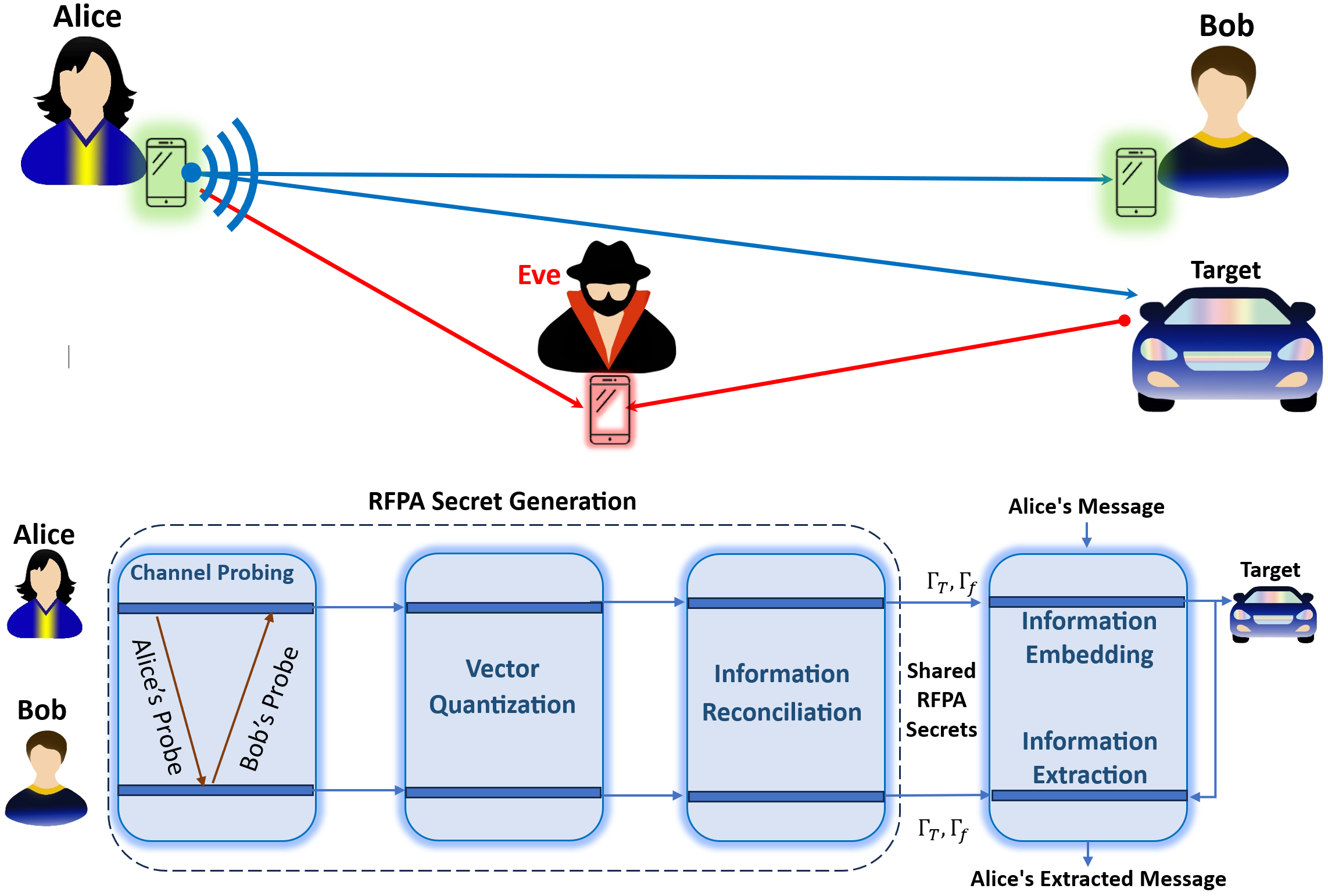}
\vspace{-1ex}
\caption{Proposed Scheme Block Diagram: Alice and Bob engage in secure \ac{RFPA}-\ac{FH}-\ac{ISAC} using channel reciprocity-based secret generation, while Eve passively eavesdrops on Alice's messages and uses the reflected echoes from the target to estimate its location and velocity.}
\label{fig:system_model}
\end{figure}

{As} depicted in Fig. \ref{fig:system_model}, {consider a scenario with} two legitimate pre-authenticated communication partners, {namely, the} \ac{ISAC} base station Alice and {a user} Bob, equipped with linear arrays of $M$ transmit and $N$ receive antennas, respectively, separated by distances $d_{T}$ and $d_{R}$.
Alice embeds information into her \ac{ISAC} \ac{FH} waveform, transmitting it towards Bob and a target (which may also be Bob), aiming to estimate the range and velocity of the target. 
Both Bob and Eve seek to exploit the embedded information received in the signal, which can be modeled as \cite{LiuSPM_2023}
\begin{align}
\label{eq:receive_signal}
\mathbf {r}(t;l) &= \mathbf {H}_l\mathbf {x}(t;l) + \mathbf {v}(t;l) \in \mathbb {C}^{N},\\
\mathbf {r}^{(e)}(t;l) &= \mathbf {H}^{(e)}_l\mathbf {x}(t;l) + \mathbf {v}^{(e)}(t;l) \in \mathbb {C}^{N},
\end{align}
where $\mathbf {H}_l\in \mathbb {C}^{N\times M}$ and $ \mathbf {H}^{(e)}_l \in \mathbb {C}^{N\times M}$ represent the flat-fading channel matrices between Alice and Bob and Alice and Eve, respectively, with elements $h_{i,j}$ and $h_{i,j}^{(e)}$ following the complex Gaussian distribution $\mathcal{CN}(0,1)$, assumed to remain constant during the $l^{\text{th}}$ \ac{FH} pulse, while $\mathbf {x}(t;l) \in \mathbb {C}^{M}$ is the transmitted \ac{ISAC} \ac{FH} signal vector at time $t$ during the $l^{\text{th}}$ \ac{FH} pulse, with $\mathbf {v}(t;l) and \mathbf {v}^{(e)}(t;l)\in \mathbb {C}^{N}$ {denoting} \ac{AWGN} at time $t$ during the $l^{\text{th}}$ \ac{FH} pulse with elements $v_{i,j}, \sim \mathcal{CN}(0,{\sigma^2_{\mathbf{v}}})$ and $v_{i,j}^{(e)}, \sim \mathcal{CN}(0,\sigma^2_{\mathbf{v}^{(e)}})${, respectively}, where ${\sigma^2_{\mathbf{v}}}$ and $\sigma^2_{\mathbf{v}^{(e)}}$ represent the power of noise at Bob's and Eve's locations, respectively. 

It is assumed hereafter that the quasi-static Rayleigh fading channel matrix $\mathbf {H}$ is perfectly known at the receiver but remains unknown {to} the transmitter. 
Eve, a passive adversary equipped with integrated sensing and communication receivers, aims to compromise the security and privacy of Alice and Bob's communication by eavesdropping and exploiting Alice's target location through knowledge of her location and estimating reflected echoes from the target.
{
Due to Eve's passive nature, it is assumed that neither legitimate partner knows Eve's location nor communication channel, thereby preventing the utilization of techniques, such as beamforming, artificial noise injection\footnotemark, or constructive interference to mitigate Eve's potential threats.}

\vspace{-1ex}

Therefore, the main objective of this paper is to leverage \ac{PLS} approaches to craft {an} \ac{ISAC} \ac{FH} waveform for the transmit signal, ensuring optimal resilience against potential Eve threats, in parallel with the improvement of radar estimation accuracy and data transmission rates.

\footnotetext{Artificial noise using random antenna selection can also be effective even without Eve’s channel knowledge, but it has limitations, including inefficient resource utilization, signal degradation, vulnerability to directional attacks, and lack of adaptability in dynamic environments \cite{DongyangVTC2017, NiuTC2022, YehTWC2021, Taneja2020, Sanayei_2004}. These limitations can be addressed by the methods proposed here, which do not exclude artificial noise techniques but rather can also be used to complement the latter to enhance system performance.}

\subsection{Signal Model for Frequency Hopping MIMO Radars}
In \ac{FH} systems, every \ac{FH} pulse lasting $\tau$ seconds comprises $Q$ sub-pulses (or chips) of duration $\Delta_t \triangleq \tau/Q$ secs, where the waveform transmitted by the $m^{\text{th}}$ antenna for a pulse can be represented as \cite{BaxterTSP_2022}
\begin{equation}
\label{eq:pulse_signal_model}
x_m(t) = \sum ^{Q-1}_{q=0} e^{{\imath
}2\pi c_{m,q}\Delta _f t }  \Pi_q(t) = \sum ^{Q-1}_{q=0}h_{m,q}(t)\Pi _q(t),
\end{equation}
where the term $h_{m,q}(t)\triangleq \exp\{{\imath}
2\pi c_{m,q} \Delta_f t\}$ represents the \ac{FH} signal transmitted by the $m^{\text{th}}$ antenna at the $q^{th}$ chip, in which $c_{m,q}$ belongs to the set of available hop codes  $\mathcal{K}\triangleq \{0, 1, \ldots, K-1\}$, and $\Pi_q(t)$ denotes the window function $\Pi(t-q\Delta_t)$, where
\begin{equation}
\label{eq:window_function}
\Pi(t)\triangleq \left\lbrace 
\begin{array}{ll}1, & 0 \leq t \leq \Delta _t,\\ 0, & {\text{otherwise}}. 
\end{array}\right.  
\end{equation}

The key design parameters for \ac{FH} waveforms in {an} \ac{ISAC} system, including $\Delta_f$, $\Delta_t$, $K$, $M${,} and $Q$, are essential for spectral confinement within the system's allocated bandwidth. 
In particular, meeting the condition $\Delta_t \triangleq 1/\Delta_f$ and ensuring $K\Delta_f \leq BW$, where $BW$ represents available radar bandwidth{, are} critical to guarantee orthogonality among hops.
The acceptable range for the number of transmit antennas $M$ is bounded by $\frac{K}{Q} \leq M \leq KQ$, when each \ac{FH} is used only once ($K = MQ$), resulting in orthogonal cross-correlation between chips and low sidelobe levels. The upper limit $M \leq KQ$ represents the maximum number of orthogonal waveforms achievable for a given bandwidth, indicating {a} high \ac{FH} recurrence rate and consequently high sidelobe levels. 
To maintain the orthogonality of \ac{FH} waveforms, each chip within the radar pulse width must satisfy the following {conditions}
\begin{equation}
\label{eq:orthogonality}
c_{m,q} \ne {c_{m^{\prime}, q}}, \quad \forall q, m \ne {m^{\prime}}.
\end{equation}

Although not obligatory for fundamental radar functionality, the condition $M \leq K$ becomes indispensable for specific information embedding techniques, guaranteeing detection by communication receivers equipped with matched filters.
%

%-------------------------------------------------
\section{Secure Frequency Hopping Waveform Design}
\label{Waveform-Design}
In this section, the existing method of embedding information into \ac{MIMO} \ac{FH} waveforms is discussed, followed by the proposal of an improved waveform version to enhance security and privacy while maintaining performance. 
Subsequently, the \ac{AF} for the proposed waveform design is calculated, demonstrating its improvement in sensing performance in \ref{subsec:AF_Evaluation_Results}.

\vspace{-0.3cm}
\subsection{State of the Art Review (\ac{FH} \ac{ISAC})}
To exemplify how \ac{SotA} information-embedding schemes can be cast into radar emissions, consider the general framework proposed in \cite{BaxterTSP_2022} and \cite{HassanienRC_2017}.
In this case,  the modulated signal on the transmitter side can be represented in the form of $M \times 1$ vector of waveforms comprising the $l^{\text{th}}$ pulse, $i.e.$,
\vspace{-1ex}
\begin{equation}
\label{eq:general_waveform}
 \mathbf{x} (t;l)\! = \!\!\!\sum ^{Q-1}_{q=0}\!\text{diag}\!\left\{{\mathbf a}_{q}^{(l)}\!\odot \!e^{{\imath}\boldsymbol{\Omega }_{q}^{(l)}}\right\} \!\exp\! \left\lbrace {\imath}2\pi {{\mathbf P}_{q}^{(l)}{\mathbf S}_{q}^{(l)}} {\mathbf d}\Delta _f t\right\rbrace \Pi_q(t),
\end{equation}
where ${\mathbf a}_{q}^{(l)}$ denotes the vector of amplitudes for the $M$ waveforms drawn from the set $\mathcal{C}_{ASK} = \{(2j-1)\Delta\mid j=1,2,\cdots, J_{ASK}\}$; $J_{ASK}$ denotes the constellation size; $\Delta$ represents the amplitude step; and $\boldsymbol{\Omega}_{q}^{(l)}$ stands for the constant phase rotations based on \ac{PSK} with constellation size of $J_{PSK}$ with the symbols ${\Omega}_{m,q}$ drawn from the constellation $\mathcal{C}_{\text{PSK}} = \left\lbrace 0,\frac{2\pi }{J},\ldots,\frac{(J-1)2\pi }{J}\right\rbrace$.

The matrix ${\mathbf P}_{q}^{(l)}$ in equation \eqref{eq:general_waveform} represents a permutation matrix of size $M \times M$, while ${\mathbf S}_{q}^{(l)}$ is a selection matrix of size $M \times K$, and $\mathbf{d} = {[0\; 1\; \cdots\; K-1]}^T$ is a vector containing the indices of all frequency hops, such that ${\mathbf c}_{q}^{(l)}={\mathbf P}_{q}^{(l)}{\mathbf S}_{q}^{(l)}\mathbf d$.
To clarify, consider the scenario where $K=6$ and $M=4$.
Then, $4$ non-iterative \ac{FH} chips can be selected from the total of $6$ available chips drawn from $\mathcal{H} \triangleq \left \{h_0, h_1, \cdots, h_5 \right\}$ for the transmit antenna array during the $q^{\text{th}}$ chip in pulse $l$.
Suppose the intention is to transmit $[h_5 \; h_3\; h_0\; h_4]^T$ which corresponds to the code vector ${\mathbf c}_{q}^{(l)} = [5 \quad 3\quad 0\quad 4]^T$. In this case, the matrix ${\mathbf S}_{q}^{(l)}$ selects the chips of interest without specific order, $[h_0 \; h_3\; h_4\; h_5]^T$ as 
\begin{subequations}
\label{eq:S_and_P}
\begin{equation}
\label{eq:selection_matrix}
\begin{aligned}
{\mathbf S}_{q}^{(l)} = 
\begin{pmatrix}
1 & 0 & 0 & 0 & 0 & 0 \\
0 & 0 & 0 & 1 & 0 & 0 \\
0 & 0 & 0 & 0 & 1 & 0 \\
0 & 0 & 0 & 0 & 0 & 1 \\
\end{pmatrix}
&
\begin{array}{c}
\rightarrow h_0 \\
\rightarrow h_3 \\
\rightarrow h_4 \\
\rightarrow h_5, \\
\end{array}
\end{aligned}
\vspace{-1ex}
\end{equation}

Then, by utilizing the permutation matrix ${\mathbf P}_{q}^{(l)}$, the order of the chips is rearranged based on the desired ${\mathbf c}_{q}^{(l)}$ for the transmit antenna array as follows.
\begin{equation}
\label{eq:permutation_matrix}
    \begin{aligned}
{\mathbf P}_{q}^{(l)} = 
\begin{pmatrix}
0 & 0 & 0 & 1  \\
0 & 1 & 0 & 0  \\
1 & 0 & 0 & 0  \\
0 & 0 & 1 & 0  \\
\end{pmatrix}
&
\begin{array}{c}
4^{\text{th}} \rightarrow 1^{\text{st}}\\
2^{\text{nd}} \rightarrow 2^{\text{nd}}\\
1^{\text{st}} \rightarrow 3^{\text{rd}}\\
3^{\text{rd}} \rightarrow 4^{\text{th}}.\\
\end{array}
\end{aligned}
\end{equation}
\end{subequations}

\vspace{-4ex}
\subsection{Proposed Generalized Secure \ac{RFPA}-\ac{FH}-\ac{ISAC} Design}
\label{RFPA-FH-ISAC}
In {the} evaluation of an \ac{AF} for waveform characteristics, conventional simple pulse trains fall short due to wide mainlobes, high sidelobes, and periodic ambiguity peaks, resulting in poor resolution and \ac{ECCM} performance \cite{EusticeWMCS_2015}.

Recent research suggests that intrapulse modulation can address these issues by narrowing the main lobe and lowering sidelobes, thereby improving range resolution and multi-target detection ability \cite{AlaeeICASSP_2019}, \cite{ZhangISJ_2016}. 
In recent times, random inter-pulse agile signals were utilized within radar systems to address ambiguity and enhance their \ac{ECCM} capabilities, which can be categorized into three types based on their agile parameters: \ac{RFA} signals, \ac{RPA} signals, and \ac{RFPA} signals.

\begin{figure}[H]
\centering
\includegraphics[width=\columnwidth]{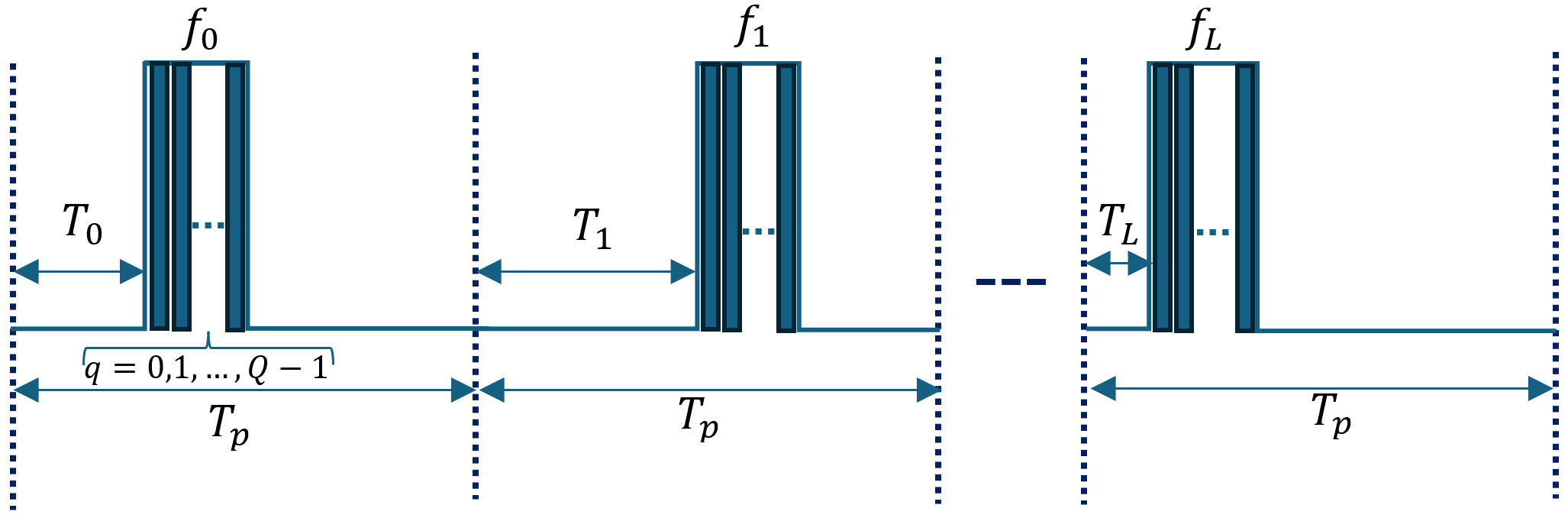}
\vspace{-3ex}
\caption{The proposed secure \ac{RFPA}-\ac{FH}-\ac{ISAC} Waveform.}
\label{fig:signal_model}
\vspace{-2ex}
\end{figure}

This approach leads to improved abilities in making unambiguous measurements and resisting clutter interference \cite{SedivyCOMITE_2013,LiuTSP_2014,XudongICR_2011}.
On the other hand, it is recognized that Eve seeks to compromise the security and privacy of communication between Alice and Bob by intercepting and exploiting the transmitted communication symbols and also {estimating} the target's location using the reflected signals.
In situations where Eve remains passive, neither legitimate partner possesses information regarding Eve's {location} or communication channel, {which hinders} the implementation of techniques to counteract Eve's potential threats.
Therefore, one of the primary {objectives} of this waveform design is to employ \ac{PLS} methods to design \ac{ISAC} \ac{FH} waveforms for the transmitted signal, ensuring optimal resilience against potential Eve's threats.
To {that end}, we utilize the inherent randomness in \ac{RFPA} signals to reduce eavesdropping risks. 
By manipulating all parameters of \ac{FH} signals, as depicted in Fig.~\ref{fig:signal_model}, we present the generalized waveform for the $m^{\text{th}}$ transmit antenna at time $t$ as
\begin{equation}
\label{eq:RFPA_waveform}
\begin{aligned}
x_m(t) = & \sum^{L-1}_{l=0}  \sum^{Q-1}_{q=0} a_{m,q}^{(l)} e^{{\imath}\Omega_{m,q}^{(l)}} e^ {{\imath}2\pi (f_l+c_{m,q}^{(l)}\Delta_f)(t-lT_p-T_l)} \\ 
&\times \Pi \left ({t-q\Delta_t-lT_p-T_l}  \right ),
\end{aligned}
\end{equation}
where $a_{m,q}^{(l)}$ and $\Omega_{m,q}^{(l)}$, respectively, represent the amplitude and fixed phase shift derived from \ac{ASK} and \ac{PSK} modulations with constellation sizes of $J_{ASK}$ and $J_{PSK}$, respectively, selected from $\mathcal{C}_{\text{ASK}}$ and $\mathcal{C}_{\text{PSK}}$ during the $q^{\text{th}}$ chip in pulse $l$.
$T_l$, random \ac{PRI} agility parameter of the $l^{\text{th}}$ pulse corresponding to the starting time of the first chip ($q=0$) in pulse $l$, falls within the range $0 \leq T_l \leq T_p - \tau$, in which $T_p$ is the PRI. $f_l$ also represents the random frequency agility parameter associated with the reference carrier frequency of the $l^\text{th}$ pulse.

The modulated {output} signal on Alice's side can be represented in the form of {an} $M \times 1$ vector corresponding to $M$ transmit antennas at each time instance \( t \) in pulse \( l \){, namely}
\vspace{-1ex}
\begin{eqnarray}
\label{eq:vector_RFPA_waveform}
\mathbf{x}(t;l) = \sum ^{Q-1}_{q=0}\text{diag}\big({\mathbf a}_{q}^{(l)}\odot e^{{\imath}\boldsymbol{\Omega }_{q}^{(l)}}\big)&& \\[-1ex]
&& \hspace{-26ex}\times \exp \left\lbrace {\imath}2\pi  (f_l \mathbf{1}_M + {\mathbf c}_{q}^{(l)} \Delta _f) (t-T_l)\right\rbrace \Pi _q(t-T_l),\nonumber 
\end{eqnarray}
where ${\mathbf a}_{q}^{(l)}$ and $\boldsymbol{\Omega}_{q}^{(l)}$ refer to the vectors of the amplitudes and the constant \ac{PSK} phase rotations associated for the $M$ waveforms, respectively.

Moreover, ${\mathbf c}_{q}^{(l)}$ is defined as the result of multiplying ${\mathbf P}_{q}^{(l)}$, ${\mathbf S}_{q}^{(l)}$, and $\mathbf d$.
In the forthcoming process, our goal is to utilize the parameters $T_l$ and $f_l$ to enhance the security of the physical layer against potential threats by Eve, achieved through a specific quantization approach as 
$T_l =  \Delta_{T_L} \times \phi_{T_l}$ and $f_l =  \Delta_{f_L} \times \phi_{f_l}.$

Let $\Delta_{T_L} \triangleq Q \Delta_t = \tau$ and $\Delta_{f_L} \triangleq K\Delta_f$ denote the constant quantization values, where $\phi_{T_l}$ and $\phi_{f_l}$ are integers randomly selected from the sets $\varphi_{T_l}\triangleq\{ 0, 1, 2, \dots, \Phi_{T_l} -1 \}$ and $\varphi_{f_l}\triangleq\{ 0, 1, 2, \dots, \Phi_{f_l} -1 \}$, respectively, and the quantities $\Phi_{T_l} \triangleq (T_p /\Delta_{T_L})-1 $ and $\Phi_{f_l} \triangleq BW/(K\Delta_f)$.
Both $\Phi_{T_l}$ and $\Phi_{f_l}$ are also designed to be powers of $2$. 
Furthermore, we define two shared secrets $\mathbf{\Gamma}_T$ and $\mathbf{\Gamma}_f$ between the legitimate partners, Alice and Bob, as $\mathbf{\Gamma}_T \triangleq \left[ \phi_{T_0}, \phi_{T_1}, \ldots, \phi_{T_{L-1}} \right]$ and $\mathbf{\Gamma}_f \triangleq \left[ \phi_{f_0}, \phi_{f_1}, \ldots, \phi_{f_{L-1}} \right]$.

After authentication, Alice and Bob receive secret vectors \(\mathbf{\Gamma}_T\) and \(\mathbf{\Gamma}_f\), which are unknown to Eve and can also be derived using the proposed \ac{CRKG} method (described in Section \ref{RFPA_Secret_Generation}).

\vspace{-2ex}
\subsection{{ Performance Analysis Based on Ambiguity Function}}
\label{subsec:AF_proposed_waveform}
In the realm of \ac{MIMO} radar signal processing, a pivotal aspect lies in the calculation and analysis of the \ac{AF}, which serves as a fundamental tool for understanding the spatial, range, and Doppler resolution characteristics influenced by the transmission of orthogonal waveforms.
{Therefore, we aim to compute the \ac{MIMO}-\ac{AF} for our \ac{RFPA}-\ac{FH}-\ac{ISAC} waveforms by utilizing the concept outlined in \cite{ChenTSP_2008}.}

Consider a target at $(\tau,\nu,f)$, where $\tau$ represents the delay associated with the target's range, $\nu$ denotes the Doppler frequency of the target, and $f$ indicates the normalized spatial frequency, defined as $f\triangleq 2\pi \frac{{d_{R}}}{\lambda} \sin \theta$, where $\theta$ denotes the angle of the target and $\lambda$ represents the wavelength.
When attempting to capture this target signal using a matched filter with assumed parameters $(\tau',\nu',f')$, the \ac{MIMO} radar \ac{AF} can be characterized as follows:
\begin{equation}
\label{eq:ambiguity_function}
\chi (\tau , \nu , f, f^{\prime}) \triangleq \sum\limits _{m=0}^{M-1}\sum\limits _{m^{\prime}=0}^{M-1} \chi_{m,m^{\prime}}(\tau , \nu) e^{{\imath}2\pi (fm-f^{\prime}m^{\prime})\gamma}, 
\end{equation}
{with} cross ambiguity function $\chi_{m,m^{\prime}}(\tau , \nu)$ {given by}
\begin{equation}
\label{eq:cross_ambiguity_function}
\chi_{m,m^{\prime}}(\tau, \nu) \triangleq \int_{-\infty}^{+\infty}x_m(t)x_{m'}^{*}(t+\tau)e^{{\imath}2\pi \nu t}{\rm d}t.
\end{equation}

To assess the sensing capabilities of the proposed \ac{RFPA}-\ac{FH}-\ac{ISAC} waveform, it is necessary to compute the MIMO radar ambiguity function.
{S}ubstituting \eqref{eq:RFPA_waveform} into \eqref{eq:cross_ambiguity_function} and considering $t_l \triangleq l T_p + T_l$ and $t_{l'} \triangleq l' T_p + T_{l'}$, we have
\begin{eqnarray}
\hspace{-5ex}\chi_{m,m^{\prime}}(\tau, \nu) = \sum_{l=0}^{L-1} \sum_{l'=0}^{L-1}\sum_{q=0}^{Q-1} \sum_{q'=0}^{Q-1} \Big(a_{m,q}^{(l)} a_{m',q'}^{(l')} e^{{\imath}\left(\Omega_{m,q}^{(l)} - \Omega_{m',q'}^{(l')}\right)}&&\hspace{-10ex}\nonumber \\
&&\hspace{-58ex} \times \int_{-\infty}^{+\infty} e^{{\imath}2\pi [(f_l+c_{m,q}^{(l)}\Delta_f)(t-t_l) - (f_{l'}+c_{m',q'}^{(l')}\Delta_f)(t+\tau-t_{l'})]}\nonumber \\[-0.5ex]
&&\hspace{-58ex}\times \Pi(t-q\Delta_t-t_l) \Pi(t+\tau-q'\Delta_t-t_{l'}) \Big) e^{{\imath}2\pi \nu t} {\rm d}t.
\label{eq:RFPA_AF(2)}
\end{eqnarray}

Replacing the variable $t$ with $t+q\Delta_t+t_l$ in \eqref{eq:RFPA_AF(2)}, we obtain 
\begin{eqnarray}
\chi_{m,m^{\prime}}(\tau, \nu) =  \sum_{l=0}^{L-1}  \sum_{l'=0}^{L-1} \sum_{q=0}^{Q-1}\sum_{q'=0}^{Q-1} a_{m,q}^{(l)} a_{m',q'}^{(l')} e^{{\imath}\left(\Omega_{m,q}^{(l)} - \Omega_{m',q'}^{(l')}\right)}&&\hspace{-12ex}\nonumber \\
&&\hspace{-58ex}\times\int_{-\infty}^{+\infty}\Big( \Pi(t) \Pi(t+(q-q')\Delta_t+t_l-t_{l'}+\tau)\nonumber \\
&&\hspace{-58ex}\times e^{{\imath}2\pi [(f_l+c_{m,q}^{(l)}\Delta_f)(t+q\Delta_t) - (f_{l'}+c_{m',q'}^{(l')}\Delta_f)(t+q\Delta_t+(t_l - t_{l'}+\tau)]}\nonumber\\
&&\hspace{-58ex}\times e^{{\imath}2\pi \nu (t+q\Delta_t+lT_p+T_n)}\Big)  {\rm d}t.
\label{eq:RFPA_AF(3)}
\end{eqnarray}
\newpage

By taking into account the lower and upper limits of the overlapping range of the window functions, defined as $\alpha_1 = \max((q'-q)\Delta_t+t_{l'}-t_l-\tau, 0)$ and $\beta_1 = \min((q'-q +1)\Delta_t+t_{l'}-t_l-\tau, \Delta_t)$, respectively, $\chi_{m, m'}(\tau, \nu)$ can be simplified to
\begin{eqnarray}
\label{eq:RFPA_AF(4)}
\chi_{m,m^{\prime}}(\tau, \nu) = \!\! \sum_{l=0}^{L-1} \sum_{l'=0}^{L-1}\sum_{q=0}^{Q-1}  \sum_{q'=0}^{Q-1} \!e^{{\imath}(\Omega_{m,q}^{(l)} - \Omega_{m',q'}^{(l')})} a_{m,q}^{(l)} a_{m',q'}^{(l')}\!&&\\[-1ex]
&&\hspace{-56ex}\times\int_{\alpha_1}^{\beta_1}\!\!\!  e^{{\imath}2\pi [(f_l+c_{m,q}^{(l)}\Delta_f)(t+q\Delta_t) - (f_{l'}+c_{m',q'}^{(l')}\Delta_f)(t+q\Delta_t+t_l-t_{l'}+\tau)]}\nonumber
\end{eqnarray}
\vspace{-5ex}
\begin{eqnarray*}
\hspace{37.5ex}\times e^{{\imath}2\pi \nu (t+q\Delta_t+t_l)}{\rm d}t&&\\
&&\hspace{-59.5ex}=\sum_{l=0}^{L-1}\sum_{l'=0}^{L-1} \sum_{q=0}^{Q-1}  \sum_{q'=0}^{Q-1} e^{{\imath}(\Omega_{m,q}^{(l)} - \Omega_{m',q'}^{(l')})} a_{m,q}^{(l)} a_{m',q'}^{(l')}\int_{\alpha_1}^{\beta_1}\!\!\!e^{\alpha_2 t +\beta_2}{\rm d}t \\[-0.5ex]
&&\hspace{-59.5ex}=\! \sum_{l=0}^{L-1}\! \sum_{l'=0}^{L-1}\! \sum_{q=0}^{Q-1}\! \sum_{q'=0}^{Q-1}\! e^{{\imath}(\Omega_{m,q}^{(l)} - \Omega_{m'\!,q'}^{(l')}\!)} a_{m,q}^{(l)} a_{m'\!,q'}^{(l')} \frac{e^{\beta_2}\!\left( e^{\alpha_2 \beta_1}\!\! - \!e^{\alpha_2 \alpha_1}\! \right)}{\alpha_2},\\
\end{eqnarray*}
where
 $\alpha_2 = {\imath}2\pi \big( (f_l+c_{m,q}^{(l)}\Delta_f) - (f_{l'}+c_{m',q'}^{(l')}\Delta_f) + \nu \big)$ and
 $\beta_2 = {\imath}2\pi \big( (f_l+c_{m,q}^{(l)}\Delta_f)q\Delta_t + \nu(q\Delta_t+t_l) (f_{l'}+c_{m',q'}^{(l')}\Delta_f)(q\Delta_t+t_l-t_{l'}+\tau)  \big).$\\
Substituting the $\alpha_2$ and $\beta_2$ into \eqref{eq:ambiguity_function}, we have
\begin{eqnarray}
\label{eq:RFPA_AF(6)}
\chi(\tau,\nu,f,f^{\prime})=\!\!\sum_{m=0}^{M-1}\!\sum_{m^{\prime}=0}^{M-1}\!\sum_{l=0}^{L-1}\! \sum_{l'=0}^{L-1}\!\sum_{q=0}^{Q-1}\!  \sum_{q'=0}^{Q-1}e^{{\imath}2\pi (mf-m^{\prime}f^{\prime})\gamma}&&\\
&&\hspace{-48ex}\quad\times\frac{e^{{\imath}(\Omega_{m,q}^{(l)} - \Omega_{m',q'}^{(l')})} a_{m,q}^{(l)} a_{m',q'}^{(l')}e^{\beta_2}\left( e^{\alpha_2 \beta_1} - e^{\alpha_2 \alpha_1} \right)}{\alpha_2}.\nonumber\hspace{-3ex}
\end{eqnarray}

The utilization of \ac{AF} proves to be a potent instrument in the examination and crafting of radar signals. {We also note that the \ac{AF} for conventional \ac{FH}-\ac{MIMO} signals can be directly derived from expressions \eqref{eq:RFPA_AF(4)} and \eqref{eq:RFPA_AF(6)} by setting the parameters \( f_l \) and \( T_l \) to zero.}
%
%======================================
\section{Information Embedding Schemes}
\label{sec:Information-Embedding-Schemes}
In the preceding section, we introduced the \ac{RFPA}-\ac{FH}-\ac{ISAC} signal model along with its \ac{AF} calculation. 
This section illustrates how these parameters can be utilized to embed information into the radar signal. 
It starts with Hybrid signaling, which boosts data rates and subsequently the complexity, and then its simplified versions, including phase-based embedding, amplitude-based embedding, and spatial index modulation, are derived depending on the intended application.
Each embedding method includes a receiver design for Bob, assuming perfect synchronization and Bob's knowledge of the frequency hops, chip interval, \ac{FH} step, and available frequency bandwidth.

\vspace{-2ex}
\subsection{Proposed Transmit Signal Design}
\label{sub:Sparse_Receiver_Hybrid}
In the \ac{ISAC} scenarios, telecommunications data is commonly integrated into radar pulses. Thus, during the \ac{PRI}s, a substantial amount of time is designated for the return of echoes from targets, making it impractical to transmit telecommunication data concurrently.
As a result, \ac{ISAC} systems frequently encounter low data transmission rates, a challenge that can be addressed by methodologies like index modulation and spatial modulation, which offer promising solutions for significantly improving data transmission rates within the pulse bandwidth.

The utilization of available frequency hops and their allocation among antenna elements facilitates the transmission of bit rate information through a combination of frequency index modulation \cite{BasarCM_2016} and spatial modulation \cite{RouTWC_2022}.
The data rates of the mentioned schemes can be enhanced by optimizing the primary radar parameters and integrating the proposed information embedding techniques with conventional modulation schemes such as \ac{QAM}. Hence, by leveraging the modified \ac{RFPA} described in \ref{RFPA-FH-ISAC}, we propose a hybrid information embedding strategy that combines multiple schemes to enhance the data rate without altering the primary radar's operating parameters.
To that end, we first define Alice's modulated signal as{}

\vspace{-0.5cm}
\begin{equation}
\label{eq:hyb_transmit_waveform}
\!\!\!\mathbf{x}_{\text{hyb}} (t;l)\! = \!\!\!\sum ^{Q-\!1}_{q=0}\!\!\text{diag}\!\left(\!{\mathbf a}_{q}^{(l)}\!\odot \!e^{{\imath}\boldsymbol{\Omega }_{q}^{(l)}}\!\right)\! e^{{\imath}2\pi {{\mathbf P}_{q}^{(l)}{\mathbf S}_{q}^{(l)}} {\mathbf d}\Delta _f t} \Pi_q(t\!-\!T_l).\!\!\!
\vspace{-1ex}
\end{equation}

In this scenario, the $M \times 1$ vectors $\boldsymbol{a}_{q}^{(l)}$ and $\boldsymbol{\Omega }_{q}^{(l)}$ consist of symbols ${a}_{m, q}^{(l)}$ and ${\Omega }_{m,q}^{(l)}$ representing the amplitudes and phases drawn from the set $\mathcal{C}_{ASK}$ and $\mathcal{C}_{PSK}$, respectively.
The matrix ${\mathbf S}_{q}^{(l)}$ selects the chips of interest (carrier frequency indices) in a non-predefined order, thereby acting as a representative of index modulation. Following this, the permutation matrix ${\mathbf P}_{q}^{(l)}$ rearranges the chip order for transmission across the transmit antennas, representing spatial modulation. Together, these operations establish the relation ${\mathbf c}_{q}^{(l)} = {\mathbf P}_{q}^{(l)}{\mathbf S}_{q}^{(l)}\mathbf d$.

In cases where prioritizing sensing accuracy over high data rates is crucial, especially when dealing with power limitations, we can effectively harness the waveform \eqref{eq:general_waveform}. 
This involves embedding information solely in the phase of the chips by employing PSK modulation, known for its superior power efficiency compared to many other modulation methods.

Therefore, the modulated signal on Alice's side can be simplified in the form of $M\times1$ vector as follows.
\begin{equation}
\label{eq:PSK_embedding}
\mathbf{x}_{\text{ph}}(t;l) \!=\! \sum ^{Q-1}_{q=0}\!\text{diag}\!\left(e^{{\imath}\boldsymbol{\Omega }_{q}^{(l)}}\right) e^{{\imath}2\pi (f_l \mathbf{1}_M + {\mathbf c}_{q}^{(l)}\Delta _f) (t-T_l)} \Pi _q(t-T_l),
\end{equation}
where the vector ${\mathbf a}_{q}^{(l)} = \mathbf{1}_M$ comprises constant amplitudes representing the $M$ waveforms, while $\boldsymbol{\Omega}_{q}^{(l)}$ denotes the fixed phase rotations corresponding to \ac{PSK}, with symbols ${\Omega}_{m,q}^{(l)}$ drawn from $\mathcal{C}_{\text{PSK}}$.

Please note that the permutation matrix ${\mathbf P}_{q}^{(l)}$ and selection matrix ${\mathbf S}_{q}^{(l)}$ are predefined and shared between Alice and Bob, conveying no additional information, thus resulting in ${\mathbf c}_{q}^{(l)}={\mathbf P}_{q}^{(l)}{\mathbf S}_{q}^{(l)}\mathbf d$.

On the other hand, in many wireless communications, \ac{ASK} might be preferred over \ac{PSK} when factors like noise resistance, ease of implementation, simplicity, or cost are crucial.
However, the choice between \ac{ASK} and \ac{PSK} depends on the specific requirements and impediments of the application. 
Both techniques can also be used simultaneously in \ac{QAM} to achieve higher data rates.
Hence, the modulated \ac{ASK} signal on Alice's side can be represented as
\begin{equation}
\label{eq:AM_embedding}
\mathbf{x}_{\text{amp}}(t;l) = \sum ^{Q-1}_{q=0}\text{diag}\{\boldsymbol{a}_{q}^{(l)}\} e^{{\imath}2\pi (f_l \mathbf{1}_M + {\mathbf c}_{q}^{(l)}\Delta _f) (t-T_l)}
 \Pi _q(t-T_l).
\end{equation}

Here, the $M \times 1$ vector $\boldsymbol{a}_{q}^{(l)}$ consists of symbols ${a}_{m, q}^{(l)}$ representing amplitudes drawn from the set $\mathcal{C}_{ASK}$,
while ${\mathbf P}_{q}^{(l)}$ and ${\mathbf S}_{q}^{(l)}$ are also predefined and shared between Alice and Bob.
Similarly, the modulated signal utilizing spatial index modulation can be further simplified as
\vspace{-0.5ex}
\begin{equation}
\label{eq:sim_embedding}
\mathbf{x}_{\text{sim}}(t;l) = \sum ^{Q-1}_{q=0} e^{{\imath}2\pi (f_l \mathbf{1}_M + {\mathbf c}_{q}^{(l)}\Delta _f) (t-T_l)} \Pi _q(t-T_l).
\vspace{-0.5ex}
\end{equation}
The matrices ${\mathbf S}_{q}^{(l)}$ and ${\mathbf P}_{q}^{(l)}$ also act as selection and permutation matrices in index modulation and spatial modulation, respectively, contributing to the relation ${\mathbf c}_{q}^{(l)}={\mathbf P}_{q}^{(l)}{\mathbf S}_{q}^{(l)}\mathbf d$.

By taking into account an \ac{AWGN} channel between Alice and Bob, the signal received by Bob can be modeled as
\vspace{-0.5ex}
\begin{equation}
\label{eq:hyb_received_signal}
\mathbf{r}_{\text{typ}}(t;l) = \mathbf{H}_l\mathbf{x}_{\text{typ}}(t;l) + \mathbf w(t;l).
\vspace{-0.5ex}
\end{equation}

Given perfect \ac{CIR} knowledge on Bob's side, he can estimate the transmitted signal as
\vspace{-0.5ex}
\begin{eqnarray}
 \hat{\mathbf{x}}_{\text{typ}}(t;l) \hspace{-4ex}&&= {\mathbf{H}_l^{\dagger}} \mathbf{r}_{\text{typ}}(t;l) 
 { \;\approx}\; \mathbf{x}_{\text{hyb}}(t;l) + {\mathbf{H}_l^{\dagger}} \mathbf w(t;l)\nonumber\\
 && =\mathbf{\Psi}_l\hat{\mathbf{s}}_{\text{type}}(t;l)+ {\mathbf{H}_l^{\dagger}} \mathbf w(t;l),
 \label{eq:estimated_transmit_signal_hyb}
\end{eqnarray}
where \( \hat{\mathbf{s}}_{\text{typ}}(t;l) \) are sparse signals \(  \hat{\mathbf{x}}_{\text{typ}}(t;l) \) projected onto the Fourier transform basis \( \mathbf{\Psi}_l \){, and} the subscript ``typ'' denotes the specific type of signal being transmitted.

Hence, there is a necessity for a receiver on the Bob side to harness information from \ac{ASK}, \ac{PSK}, Index, and Spatial modulations by estimating $\hat{\mathbf a}_{q}^{(l)}$, $\hat{\boldsymbol{\Omega}}_{q}^{(l)}$, $\hat{\mathbf S}_{q}^{(l)}$ and $\hat{\mathbf P}_{q}^{(l)}$, respectively, from $\hat{\mathbf{x}}_{\text{hyb}}(t;l)$.\\
\vspace{-2em}
\subsection{Proposed Receiver Design}
To extract the inherent symbols encoded when a \ac{PSK} signal \(\mathbf{x}_{\text{ph}}(t;l)\) is transmitted, matched filtering serves as an optimal linear filtering technique designed to maximize the \ac{SNR} in the presence of additive stochastic noise.
Hence, the vector of $K$ available \ac{FH} waveforms for pulse $l$ for matched filtering is defined as
\begin{eqnarray}
\label{eq:available_FH_waveforms_PSK}
\mathbf {h}(t;l) = e^{{\imath}2\pi (f_l \mathbf{1}_K + \textbf {d}\Delta _f (t-T_l))}&& \\ 
&&\hspace{-26ex} = e^{{\imath}2\pi\left[f_l (t-T_l),\;  (f_l +\Delta _f) (t-T_l),\; \dots \;,\;  (f_l + (K-1)\Delta _f) (t-T_l)\right]^T}\!\!. \nonumber
\end{eqnarray} 

Given the assumption that Bob possesses knowledge of Alice's \ac{FH} sequence or the same ${\mathbf P}_{q}^{(l)}$ and ${\mathbf S}_{q}^{(l)}$ at each chip $q$ in pulse $l$, he can compute the vectors of transmitted hops  as $\tilde{\mathbf h}_q(t;l)\triangleq {\mathbf P}_{q}^{(l)}{\mathbf S}_{q}^{(l)} \mathbf h(t;l) =[\tilde{h}_{0,q}(t;l),\tilde{h}_{1,q}(t;l)\ldots\tilde{h}_{M-1,q}(t;l)]^T$.
Applying matched filtering to the \ac{FH} chips results in
\begin{equation}
\label{eq:matched_filtering_PSK}
 \mathbf{\gamma}^{(l)}_{q} = \int _{T_l + q\Delta _t}^{T_l + (q+1)\Delta _t}\left(\mathbf{1}_M\cdot\hat{\mathbf{x}}_{\text{ph}}(t;l)\right)\tilde{\mathbf h}_{q}^{*}(t;l){\rm d}t,  
\end{equation} 

Then, the embedded phases in the $q^{\text{th}}$ chip in pulse $l$, can be exploited as the phase of the estimated symbols as
\begin{equation}
\label{eq:PSK_symbols_estimation_PSK}
\hat{\boldsymbol{\Omega}}_{q}^{(l)} = \angle {\boldsymbol{{\gamma}}_{q}^{(l)}}.
\end{equation} 

Similarly, when an \ac{ASK} signal \(\mathbf{x}_{\text{amp}}(t;l)\) is transmitted, Bob can apply matched filtering to the \ac{FH} chips, allowing the exploitation of the embedded amplitude information, thus
\begin{equation}
\label{eq:matched_filtering_AM}
\hat{\mathbf a}_{q}^{(l)} = \int _{T_l + q\Delta _t}^{T_l + (q+1)\Delta _t}
\left(\mathbf{1}_M \cdot \hat{\mathbf x}_{\text{amp}}(t;l)\right)\tilde{\mathbf h}_{q}^{*}(t;l){\rm d}t.  
\end{equation} 

\begin{algorithm}[H]
\caption{Sparse Receiver Design for Spatial Index Mod.}
\label{alg:Sim_Sparse_Receiver}
\scriptsize
\begin{algorithmic}[1]
\renewcommand{\algorithmicrequire}{\textbf{Input:}}
\renewcommand{\algorithmicensure}{\textbf{Output:}}
\REQUIRE $\mathbf{r}_{\text{sim}}(t;l)$ and $ \mathbf{H}_l$
\ENSURE  ${\mathbf S}_{q}^{(l)}$ and ${\mathbf P}_{q}^{(l)}$
\vspace{0.3ex}
\hrule
\vspace{0.3ex}
\hrule
\vspace{0.5ex}
\FOR{each pulse $l$}
\FOR{each sub-pulse $q$} 
    \STATE Calculate $\hat{\mathbf{x}}_{\text{sim}}(t;l) = {\mathbf{H}_l^{\dagger}} \mathbf{r}_{\text{sim}}(t;l)$.
    \FOR{each antenna element $m$}
        \STATE $\rho = (\hat{s}_{m,q}(:;l))$
        \STATE $\ell_{\rho} = \text{length}(\rho)$.
        \STATE Let $\mathbf{\Psi}(i,j) \triangleq \ \exp\{-{\imath} 2 \pi i j / \ell_{\rho}\}, \forall i, j = 0, \dots, \ell_{\rho}-1.$
        \STATE Select atom: $\hat{c}_{m,q}^{(l)} = (\arg\max_i |\langle \mathbf{\Psi}_i, \rho \rangle|)\times \frac{f_s}{\ell_{\rho}}$.
    \ENDFOR
    \STATE Compute $\hat{{\mathbf S}}_{q}^{(l)}$ and $\hat{{\mathbf P}}_{q}^{(l)}$ so that $\hat{{\mathbf c}}_{q}^{(l)}=\hat{{\mathbf P}}_{q}^{(l)}\hat{{\mathbf S}}_{q}^{(l)}\mathbf d$ via \eqref{eq:S_and_P}.
\ENDFOR
\ENDFOR
\end{algorithmic} 
\end{algorithm}
\vspace{-1ex}
\begin{algorithm}[H]
\caption{Sparse MF Receiver Design for Hybrid Mod.}
\label{alg:Sparse_Matched_Filter_Receiver}
\scriptsize
\begin{algorithmic}[1]
\renewcommand{\algorithmicrequire}{\textbf{Input:}}
\renewcommand{\algorithmicensure}{\textbf{Output:}}
\REQUIRE $\mathbf{r}_{\text{hyb}}(t;l)$ and $\mathbf{H}_l$
\ENSURE  $\hat{\mathbf a}_{q}^{(l)}$, $\hat{\boldsymbol{\Omega}}_{q}^{(l)}$, $\hat{\mathbf S}_{q}^{(l)}$ and $\hat{\mathbf P}_{q}^{(l)}$
\FOR {each pulse $l=0$ to $L-1$}
\STATE Calculate $\mathbf {h}(t;l)$ based on \eqref{eq:available_FH_waveforms_PSK}.
\FOR{each sub-pulse $q=0$ to $Q-1$}
    \STATE Calculate $\hat{\mathbf{x}}_{\text{hyb}}(t;l) = {\mathbf{H}_l^{\dagger}} \mathbf{r}_{\text{hyb}}(t;l)$.
    \FOR{each antenna element $m=0$ to $M-1$}
        \STATE $\rho = (\hat{s}_{m,q}(:;l))$
        \STATE $\ell_{\rho} = \text{length}(\rho)$.
        \STATE Let
        $\mathbf{\Psi}(i,j) \triangleq \exp\{-{\imath} 2 \pi i j / \ell_{\rho}\},  \forall i, j = 0, \dots, \ell_{\rho}-1.$
        \STATE Select atom: $\hat{c}_{m,q}^{(l)} = (\arg\max_i |\langle \mathbf{\Psi}_i, \rho \rangle|)\times \frac{f_s}{\ell_{\rho}}$.
    \ENDFOR
    \STATE Form $\hat{\mathbf{c}}_{q}^{(l)}=\big[\hat{c}_{0,q}^{(l)}, \hat{c}_{1,q}^{(l)}, \cdots, \hat{c}_{M-1,q}^{(l)}\big]$
    \STATE Compute $\hat{{\mathbf S}}_{q}^{(l)}$ and $\hat{{\mathbf P}}_{q}^{(l)}$ so that $\hat{{\mathbf c}}_{q}^{(l)}=\hat{{\mathbf P}}_{q}^{(l)}\hat{{\mathbf S}}_{q}^{(l)}\mathbf d$ via \eqref{eq:S_and_P}
    \STATE Define $\tilde{\mathbf h}_q(t;l)\triangleq \hat{{\mathbf P}}_{q}^{(l)} \hat{{\mathbf S}}_{q}^{(l)} \mathbf h(t;l)$
    \STATE Using Matched filtering \\
    $\boldsymbol{\gamma}^{(l)}_{q} = \int _{T_l + q\Delta _t}^{T_l + (q+1)\Delta _t} \left(\mathbf{1}_M\cdot\hat{\mathbf{x}}_{\text{hyb}}(t;l)\right)\tilde{\mathbf h}_{q}^{*}(t;l){\rm d}t$,
    \STATE $\hat{\mathbf a}_{q}^{(l)} = \mid{\boldsymbol{\gamma} ^{(l)}_{q}}\mid$ and $\hat{\boldsymbol{\Omega}}_{q}^{(l)} = \angle{\boldsymbol{\gamma} ^{(l)}_{q}}$
\ENDFOR
\ENDFOR
\end{algorithmic} 
\end{algorithm}

In \ac{SIM}, extracting the information embedded in \({\mathbf S}_{q}^{(l)}\) and \({\mathbf P}_{q}^{(l)}\) requires more computational effort. 
Therefore, Bob can employ the optimal \ac{ML} receiver for the \ac{AWGN} channel to efficiently process the signals, {namely}
\begin{equation}
\label{eq:ML}
\left\lbrace \hat{\mathbf{c}}_q^{(l)}\right\rbrace _{k=0}^{K-1} = \mathop {\arg \min } \limits _{\lbrace{\mathbf{c}}_q^{(l)}\rbrace } \big\Vert \boldsymbol{\mathbf{x}}_{\text{sim},q}(t;l) - \hat{\mathbf{x}}_{\text{sim},q}(t;l)\big\Vert _2^2.
\end{equation}

Notice that solving this optimization problem requires exploring the entire space of possible combinations of \({\mathbf S}_{q}^{(l)}\) and \({\mathbf P}_{q}^{(l)}\), leading to high computational complexity, as typically discussed in the literature \cite{BaxterTSP_2022, HassanienRC_2017}. 
However, recognizing the sparse nature of the received signal in the frequency domain, we propose a receiver architecture with reduced complexity to address the computational challenges associated with spatial index modulation decoding.
Given the sparsity of the transmit signal in the frequency domain, we define the dictionary matrix \( \mathbf{\Psi}_l \) as the local Fourier transform basis, which yields an \( M \)-sparse signal for each chip \( q \) across all \( M \) transmit antennas in pulse \( l \), or a \( 1 \)-sparse signal for each chip \( q \) and transmit antenna \( m \), so that \eqref{eq:ML} can be rewritten as
\begin{equation}
\label{eq:sparse_ML}
\begin{aligned}
\left\lbrace \hat{\mathbf{c}}_q^{(l)}\right\rbrace _{k=0}^{K-1} = & \mathop {\arg \min } \limits _{\lbrace{\mathbf{c}}_q^{(l)}\rbrace } \big\Vert \boldsymbol{\mathbf{x}}_{\text{sim},q}(t;l) - \mathbf{\Psi}_l\hat{\mathbf{s}}_{\text{sim},q}(t;l)\big\Vert _2^2\\
& \quad\quad \mathrm{s.t.}\; \big\Vert \hat{\mathbf{s}}_{\text{sim},q}(t;l) \big\Vert _{l_1}=M.
\end{aligned}
\end{equation}

Since the signal by Bob is $1$-sparse in the frequency domain for the $q^{\text{th}}$ subpulse in pulse $l$, the \ac{OMP}, described in Algorithm \ref{alg:Sim_Sparse_Receiver}, is used \cite{DavenportTIT_2010,CaiTIT_2011}.

Widely employed due to its simplicity and effectiveness, \ac{OMP} efficiently recovers sparse signals from limited measurements, making it suitable for blind frequency hopping recovery tasks.
Finally, we propose a low-complexity sparse receiver design to extract the information embedded in \({\mathbf S}_{q}^{(l)}\) and \({\mathbf P}_{q}^{(l)}\), as outlined in Algorithm \ref{alg:Sim_Sparse_Receiver}.
When a hybrid modulation scheme is employed, the receiver on Bob's side must extract information from \ac{ASK}, \ac{PSK}, Index, and Spatial modulations by estimating \(\hat{\mathbf a}_{q}^{(l)}\), \(\hat{\boldsymbol{\Omega}}_{q}^{(l)}\), \(\hat{\mathbf S}_{q}^{(l)}\), and \(\hat{\mathbf P}_{q}^{(l)}\), respectively, from the received signal \(\hat{\mathbf{x}}_{\text{hyb}}(t;l)\).

To achieve this, a sparse receiver on Bob’s side is proposed using \ac{OMP} and matched filtering, enabling efficient extraction of information from the noisy hybrid-modulated {(hereafter termed HYB)} signal, as detailed in Algorithm \ref{alg:Sparse_Matched_Filter_Receiver}.

\vspace{-2ex}
\subsection{Balancing Parameter Trade-offs in System Design}
We aim to enhance security and radar performance in \ac{ISAC} scenarios by using \ac{PRI} and frequency agility. 
{But \ac{PRI} agility requires precise synchronization, while frequency agility reduces transmission bandwidth and bit rate.
Adjusting the frequency agility parameter ($\Phi_{f_l}$) allows a trade-off between data rate, target velocity estimation, frequency synchronization, and privacy. 
For high radar accuracy and security, increasing $\Phi_{f_l}$ is beneficial, whereas for maximizing data rates, lowering it or setting it to 1 is better, so that \ac{PRI} agility can still be used effectively, and if synchronization challenges occur, reducing or setting $\Phi_{T_l}$ to 1 can help achieve synchronization at the cost of losing \ac{PRI} agility.}

Security risks during sequence assignment arise from potential leakage of preshared values of $T_l$ and $f_L$ between authenticated partners, but this risk can be mitigated using our proposed channel reciprocity-based shared sequence techniques in low-interference scenarios.

%===============================
\vspace{-2ex}
\section{RFPA Secret Generation}
\label{RFPA_Secret_Generation}

In this section, a novel approach is introduced to enhance data security through the generation and utilization of shared secrets in communication systems, leveraging the \ac{CIR} of \ac{MIMO} channels in \ac{FH} techniques in an innovative fashion, the method ensures maximum entropy and randomness in shared secret generation, effectively mitigating the risks of eavesdropping by providing a robust foundation for secure transmission.

By implementing real-time secret generation, the approach minimizes eavesdropping risks, enhances information privacy, and prevents attackers from exploiting carrier frequencies or estimating target locations, thereby reinforcing overall system security and privacy.
These protocols unfold through several stages: channel Probing, where both parties observe correlated samples from a common randomness source of the wireless \ac{CIR}; 
quantization, which converts these samples into shared symbols; 
information reconciliation, which corrects any discrepancies between Alice's and Bob's observed and quantized binary sequences;
{and} privacy amplification, which mitigates information leakage to Eve during earlier stages \footnotemark.
\vspace{-1ex}

\footnotetext{{\ac{FH} sequence optimization is also a technique that can enhance sensing performance and anti-jamming capabilities. Future research should explore the feasibility of collaborative implementation between Alice and Bob while ensuring non-correlation with Eve’s optimizers to strengthen security.}}

In this paper, the first three steps generate shared secrets for Alice and Bob.
Unlike other approaches that use these secrets solely for encryption, this work focuses on enhancing \ac{PLS} and \ac{RFPA} through these physical layer key generation. { The final shared secrets, \(\mathbf{\Gamma}_T\) and \(\mathbf{\Gamma}_f\), serve a dual purpose as follows: they obfuscate both the Doppler frequency and pulse start times, significantly complicating passive adversaries' attempts to estimate the target's velocity and range, respectively, while also enabling Alice and Bob to maintain their frequency and \ac{PRI} synchronization across \(L\) \ac{FH} pulses.}

\vspace{-2ex}
\subsection{Channel Probing}
When initiating the establishment of a shared secret over a wireless fading channel between Alice and Bob, the essential first step involves bi-directional channel probing, vital for key generation in wireless communication.
They utilize \ac{TDD} systems, employing a single pre-defined carrier frequency \(f_0\) for both directions, thereby ensuring a stable channel status throughout the coherence time (\(T_{coh}\)). 
Alice initiates transmission by sending a request pilot signal to Bob, prompting him to estimate \(\mathcal B\) as {a} randomness source, defined as the summation of signals received by various receive antennas.
The \ac{CIR} integrates multi-path components, each characterized by attenuation \(\alpha_{l}(t)\), phase shift \(\psi_{l}(t)\), and delay \(\tau_{l}(t)\) for the \(l^{th}\) path, serving as essential inputs for \ac{CRKG}. This relationship is mathematically described by
\vspace{-1ex}
\begin{equation}
\label{eq:channel-model}
\mathcal B = \sum_{l=1}^{L} \alpha_{l}(t) e^{{\imath}\psi_{l}(t)} \delta(t-\tau_{l}(t)).
\end{equation}

After a brief delay, Bob acknowledges by transmitting his pilot signal, enabling Alice to similarly measure $\mathcal A$, a reciprocal counterpart to $\mathcal B$.
It is assumed herein that the collection of at least $L$ significant samples of $\mathcal A$ and $\mathcal B$ is feasible, denoted as $\mathcal A_l$ and $\mathcal B_l$, where $L$ aligns with the number of pulses transmitted within a coding period.
This period repeats twice: once for the assignment of $T_l$ and once for the assignment of $f_l$.
It should be noted that Eve is assumed to be sufficiently distant from Alice and Bob, such that her channel randomness source $\mathcal E$ is non-reciprocal with $\mathcal A$ and $\mathcal B$.
To enhance security during the upcoming coding phase, it is advisable to carry out this process using the last carrier frequency $f_L$ employed in the current period, which reduces the likelihood of detection by Eve, who may be unaware of the current secret assignment.

\vspace{-2ex}
\subsection{Vector Quantization and Information Reconciliation}
\label{Vector Quantization Design}
In this approach, the novel \ac{VQ} algorithm \ref{alg:modified-FCmeans} is proposed that introduces a set of quantization symbols to the randomness sources $\mathcal A$, $\mathcal B$, and $\mathcal E$.

The primary aim is to achieve a uniform distribution of symbol sequences within the quantization symbol sets $\varphi_{T_l}$ and $\varphi_{f_l}$, ensuring equal probabilities for each symbol in the set and consequently enhancing the entropy of the system.
This uniformity increases the entropy, thereby enhancing the security of the generated secrets against guessing attempts by Eve.
The \ac{FCM} algorithm has been modified to efficiently associate each random value with a specific symbol, ensuring the creation of clusters of equal size, which results in maximum entropy and randomness of output secret symbols \cite{BagheriGIIS_2024}.

In Phase I, the \ac{FCM} algorithm is utilized, a widely recognized clustering technique that relies on membership degrees to express the level of connection between data points and clusters.
This algorithm enables a soft assignment of data points to multiple clusters by minimizing an objective function and measuring the weighted distance between data points and cluster centers.

The membership probabilities, denoted as $u_{l,\phi_{T_l}}$ and $u_{l,\phi_{f_l}}$ for CIR samples $\mathcal{A}_l$ and $\mathcal{B}_l$ respectively, represent the degree of belongingness to clusters $\phi_{T_l}$ and $\phi_{f_l}$, which can be computed using the following equations
\begin{eqnarray}
\label{eq:membership_probability_T_l}
&u_{\mathcal A_l,\phi_{T_l}} = \Big( \sum_{\phi=0}^{\Phi_{T_l}-1} \big( \frac{d_{\mathcal A_l,\phi_{T_l}}}{d_{\mathcal A_l,\phi}} \big)^{\frac{2}{m-1}} \Big)^{\!\!-1}\!\!\!\!\!,&\\ 
\label{eq:membership_probability_f_l}
& u_{\mathcal A_l,\phi_{f_l}} = \Big( \sum_{\phi=0}^{\Phi_{f_l}-1} \big( \frac{d_{\mathcal A_l,\phi_{f_l}}}{d_{\mathcal A_l,\phi}} \big)^{\frac{2}{m-1}} \Big)^{\!\!-1}\!\!\!\!\!,&
\end{eqnarray}
where $\Phi_{T_l}$ and $\Phi_{f_l}$ are the total number of clusters; $d_{\mathcal A_l,\phi}$ is the distance between CIR sample $\mathcal A_l$ and cluster center $\phi$, and $m$ is a parameter controlling the fuzziness of the clustering.  

The cluster centers $\upsilon_{\phi_{T_l}}$ and $\upsilon_{\phi_{f_l}}$ are computed as the weighted average of data points, namely,

\quad\\[-4ex]
\begin{equation}
\label{eq:cluster_centers_T}
\upsilon_{\phi_{T_l}} = \frac{\sum\limits_{\phi=0}^{\Phi_{T_l}-1} u_{\mathcal A_l,\phi_{T_l}} ^m \cdot \mathcal A_l}{\sum\limits_{\phi=0}^{\Phi_{T_l}-1} u_{\mathcal A_l,\phi_{T_l}}^m}\: \text{and}\:
 \upsilon_{\phi_{f_l}} = \frac{\sum\limits_{\phi=0}^{\Phi_{f_l}-1} u_{\mathcal A_l,\phi_{f_l}} ^m \cdot \mathcal A_l}{\sum\limits_{\phi=0}^{\Phi_{f_l}-1} u_{\mathcal A_l,\phi_{f_l}}^m}.
\end{equation}

These formulas provide the essential mathematical framework for implementing \ac{FCM} and obtaining membership probabilities and cluster centers in the clustering process. 

In phase II, each data point's likelihood of belonging to a cluster is represented by \( u_{\mathcal A_l,\phi_{T_l}}\); upon finding the cluster $\phi^*_{T_l}$ below the desired size (\(L/\Phi_{T_l}\)), the algorithm assigns the data point $\mathcal{A}^*_l$ with the maximum belongingness $u_{\mathcal{A}^*_l,\phi^*_{T_l}}$ to it, followed by setting $u_{\mathcal{A}^*_l,\phi^*_{T_l}}$ to $0$ to prevent the point from being assigned to multiple clusters.
If the cluster surpasses the desired size, $u_{\mathcal{A}^*_l,\phi^*_{T_l}}$ is set to \(0\), {to} halt further expansion, ensuring balanced cluster sizes.
Existing studies {indicate} that following \ac{VQ}, Alice and Bob engage in exchanging cluster centers via the wireless channel to align their cluster labels \cite{HanIS_2020, HongTIFS_2017, ChenTWC_2022}. 
However, this methodology introduces security vulnerabilities, as it is susceptible to eavesdropping, and results in heightened communication overhead and delays in the establishment of the shared secret key.

In Phase III, the transmission of cluster centers is efficiently mitigated, thereby substantially reducing complexity by eliminating the need for their exchange. 
Instead, Alice and Bob opt for a simplified approach, assigning cluster labels through a direct numbering scheme ranging from $0$ to $\Phi_{T_l}-1$.
This assignment strategy is grounded on equalizing the distribution of data points across each cluster.

In particular, the distance matrices \(\mathbf{D}_x\) and \(\mathbf{D}_y\) for the real and imaginary parts of the central values are calculated in Phase III, with subsequent matrices initialized accordingly.
Close distance thresholds \(t_{x_\upsilon}\) and \(t_{y_\upsilon}\) based on standard deviations, sorts the centers in ascending order of \(x\) are also established, and then iterative updates of the centers' numbers are obtained.

\begin{algorithm}[H]
\scriptsize
\caption{The proposed VQ for FH secret generation}
\label{alg:modified-FCmeans}

\begin{algorithmic}[1]
\renewcommand{\algorithmicrequire}{\textbf{Input:}}
\renewcommand{\algorithmicensure}{\textbf{Output:}}
\REQUIRE Data array $\mathcal A_l=(x_{\mathcal A_l}, y_{\mathcal A_l})$,
        $l=0, 1, \dots ,L-1$;
        $\Phi_{T_l}$ random cluster centers $\upsilon_{\phi_{T_l}}$ for PRI agility; 
        Iteration step $t=0$, convergence threshold $\epsilon$.
        $z$ is a scale parameter of Phase III.
\ENSURE  $\Phi_{T_l}$ clusters $G$ with $L/\Phi_{T_l}$ data points
\vspace{0.3ex}
\hrule
\vspace{0.3ex}
\hrule
\vspace{0.75ex}
\hspace{-3.6ex}\textbf{Phase I: Fuzzy C-Means clustering (FCM)}\\
%\vspace{0.5\baselineskip}
\WHILE{$\parallel \pi_{{\Phi}}^{t} - \pi_{{\Phi}}^{t-1}\parallel \leq \epsilon$,}
%\vspace{0.5\baselineskip}
\FOR{each $\mathcal A_l \in \mathcal A$ and $\phi_{T_l} \in [0, \Phi_{T_l}-1]$,}
    \STATE Calculate $u_{\mathcal A_l,\phi_{T_l}}$ according to Eq. \ref{eq:membership_probability_T_l}.
\ENDFOR
%\vspace{0.5\baselineskip}
%
%\vspace{0.5\baselineskip}
\FOR{each $\phi_{T_l} \in [0, \Phi_{T_l}-1]$,}
    \STATE Calculates the new centers based on Eq. \eqref{eq:cluster_centers_T}.
\ENDFOR
%\vspace{0.5\baselineskip}
%
\STATE $t += 1$, and calculate the new partition matrix $\pi_{\mathcal{\phi}}^{t}=[u_{\mathcal A_l,\phi_{T_l}}]$.
\ENDWHILE
\vspace{0.5ex}
\hrule
\vspace{0.75ex}
\hspace{-4.2ex}
\textbf{Phase II: Equalizing the size of clusters}\\
\vspace{0.5ex}
\STATE Initialize empty clusters $G_0, G_1, \dots, G_{\Phi_{T_l}-1}$.
\vspace{0.5\baselineskip}
\WHILE{$\mid G_0 \mid= \mid G_1\mid = \dots = \mid G_{\Phi_{T_l}-1}\mid = L/\Phi_{T_l}$}
\STATE Calculate $\mathcal{A}^*_l,\phi^*_{T_l} = \underset{\substack{\mathcal A_l,\phi_{T_l}}}{\arg\max}\ u_{\mathcal A_l,\phi_{T_l}}$
%
%\vspace{0.5\baselineskip}
\IF{$\mid G_{\phi^*_{T_l}}\mid \leq L/{\Phi_{T_l}}$}
    \STATE Assign $\mathcal{A}^*_l$ to cluster $G_{\phi^*_{T_l}}$ and replace $u_{\mathcal{A}^*_l,\phi^*_{T_l}}$ with $0$
\ELSE
    \STATE Replace $u_{\mathcal{A}^*_l},\phi^*_{T_l}$ with 0
\ENDIF
%\vspace{0.5\baselineskip}
\ENDWHILE
%\vspace{0.5\baselineskip}
%
\STATE Update the new centers based on Eq. \eqref{eq:cluster_centers_T}
\vspace{0.5ex}
\hrule
\vspace{0.75ex}
\hspace{-4.3ex}
\textbf{Phase III: Equalizing the labels on both sides of Alice and Bob}\\
\vspace{0.5ex}
\STATE Standardize features \(x_{\upsilon}\) and \(y_{\upsilon}\) of centers by removing the mean and scaling to unit variance (${\upsilon}_{\phi_{T_l}}=(x_{\upsilon_{\phi_{T_l}}}, y_{\upsilon_{\phi_{T_l}}})$).  
\STATE Compute square distance matrices \(\mathbf{D}_{x_{\upsilon}}\) and \(\mathbf{D}_{y_{\upsilon}}\) for centers using \(\mathbf{D}_{x_{\upsilon}}[i, j] = x_{{\upsilon}_i} - x_{{\upsilon}_j}\) and $\mathbf{D}_{y_{\upsilon}}[i, j] =y_{\upsilon_i} - y_{\upsilon_j}$.
\STATE Calculate the standard deviation of elements in \(\mathbf{D}_{x_{\upsilon}}\) and \(\mathbf{D}_{y_{\upsilon}}\) as \(\sigma_{x_{\upsilon}}\) and \(\sigma_{y_{\upsilon}}\).
\STATE Set close distance thresholds \(t_{x_{\upsilon}} = \sigma_{x_{\upsilon}} / z\) and \(t_{y_{\upsilon}} = \sigma_{y_{\upsilon}} / z\).
\STATE Number centers from $1$ to \({\Phi_{T_l}}\) based on the \(x_{\upsilon}\) dimension, in ascending order.
%\vspace{0.5\baselineskip}
\WHILE{the centers remain unchanged,}
%\vspace{0.5\baselineskip}
\FOR{each cluster center $\phi \in [0, \Phi_{T_l}-1]$,}
    %\vspace{0.5\baselineskip}
    \IF{\((x_{{\upsilon}_{\phi}} - x_{{\upsilon}_{{\phi}+1}}) < t_{x_{\upsilon}}\) and \((y_{{\upsilon}_{\phi}} - y_{{\upsilon}_{{\phi}+1}}) \geq t_{y_{\upsilon}}\),}
        \STATE Swap the numbering of centers \(\phi\) and \({\phi}+1\).
    \ENDIF
    %\vspace{0.5\baselineskip}
    %
    \IF{\((x_{{\upsilon}_{\phi}} - x_{{\upsilon}_{{\phi}+1}}) < t_{x_{\upsilon}}\) and \((y_{{\upsilon}_{\phi}} - y_{{\upsilon}_{{\phi}+1}}) < t_{y_{\upsilon}}\),}
        \STATE Number centers based on \(x_{\upsilon_{\upsilon}} + y_{\upsilon_{\upsilon}}\).
    \ENDIF
    %\vspace{0.5\baselineskip}
\ENDFOR
%\vspace{0.5\baselineskip}
\ENDWHILE
%\vspace{0.5\baselineskip}
\STATE Update the labels $G_{\phi_{T_l}}$ based on the new centers numbering.
\end{algorithmic} 
\end{algorithm}
\vspace{-2ex}

During each of such iterations, it compares the \(x\) and \(y\) values of two neighboring centers, and if the difference in their \(x\) values is smaller than \(t_{x_\upsilon}\) and the difference in their \(y\) values is larger than \(t_{y_\upsilon}\), it swaps their centers numbers. 
Otherwise, it renumbers the centers in ascending order based on their \(x+y\) values.
The iterative process persists until centers remain unaltered, indicating convergence, with the algorithm generating shared secrets for \ac{PRI} agility, such that a repetition of the latter {is} required {to} establish shared secrets between Alice and Bob for frequency agility.
\vspace{-3ex}
\begin{figure}[H]
\centering
\includegraphics[width=1\columnwidth]{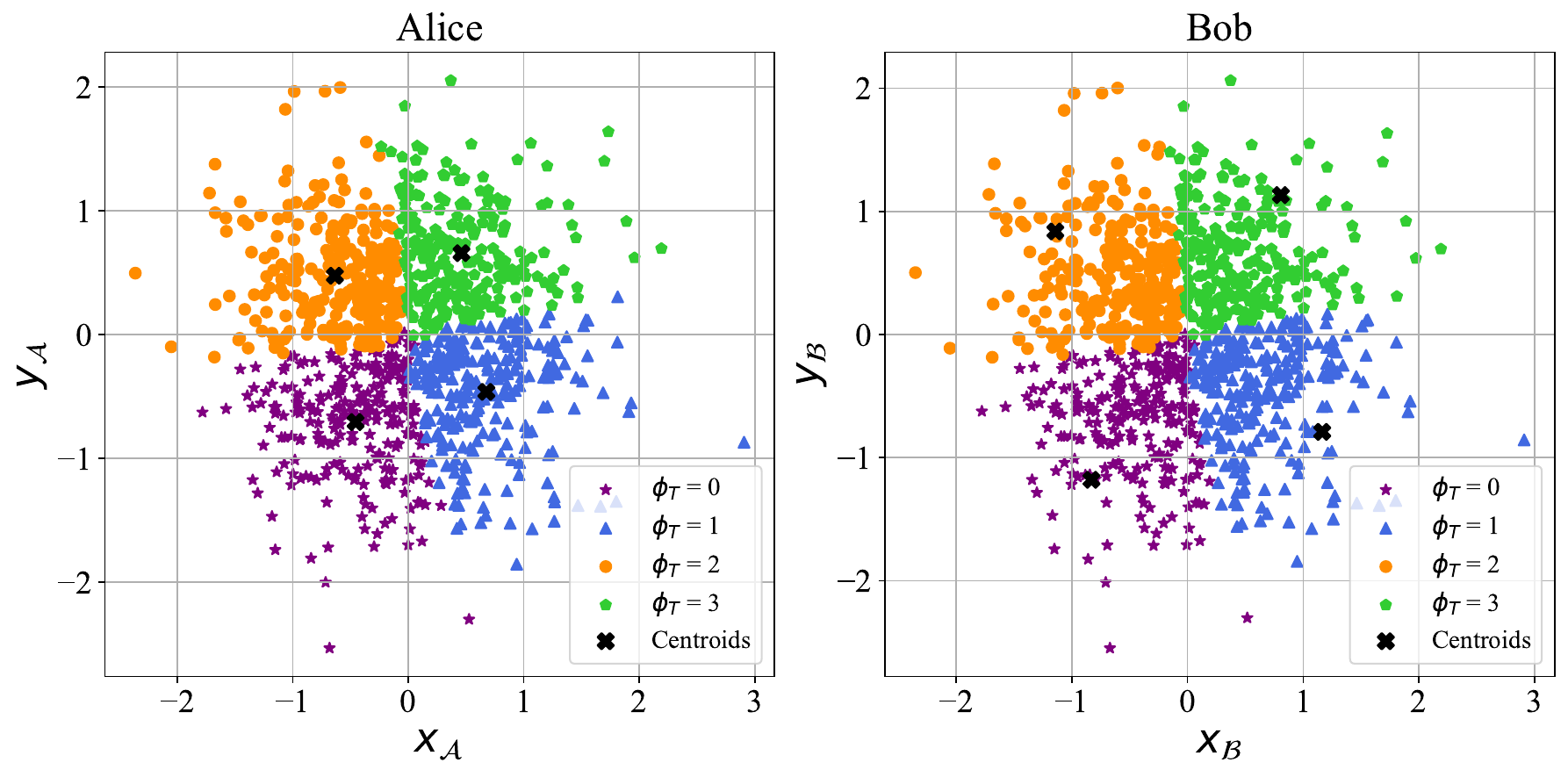}
\vspace{-4ex}
\caption{An illustration of the grouped sample points for Alice and Bob as determined by the proposed algorithm, using either $\Phi_{T_l} = 4$ or $\Phi_{f_l} = 4$.}
\label{fig:CRKG}
\vspace{-1ex}
\end{figure}

Fig. \ref{fig:CRKG} shows the clustering results from the proposed algorithm, featuring 4 clusters and 1024 CIR sample points for Alice and Bob following the quantization process. 
Each cluster contains an equal distribution of 256 samples, assigned based on the minimum distance to the cluster center.
Despite generating synchronized sequences for $f_l$ and $T_l$, noise and imperfect channel reciprocity can cause mismatches between Alice and Bob's sequences, necessitating equalization to avoid high \ac{BER}, which is discussed in the sequel.

\noindent \emph{Note: Information Reconciliation Scheme}

Information reconciliation schemes typically involve multiple exchanges to resolve inconsistencies in key bits, unsuitable for 6G's low latency requirements. Error correction code-based approaches like Polar Codes facilitate reconciliation between Alice and Bob's bitstrings to ensure identical secret keys. The process involves converting channel samples to bitstrings, generating a random number vector, applying CRC and polar encoding, rate matching, and XOR operations to reconcile sequences, resulting in both parties having the same secret key. This method, detailed in \cite{PengICACES_2022, TalTIT_2015, ShakibaSSP_2021}, enhances the reliability and security of the information reconciliation process.
%============================

\vspace{-1ex}
\section{Complexity Analysis}
\label{Complexity}
Analyzing the computational complexity of the Sparse Matched Filter Receiver algorithm provides insights into the complexity {of both} Algorithms \ref{alg:Sparse_Matched_Filter_Receiver} and \ref{alg:Sim_Sparse_Receiver}, as the former includes the latter's steps.
The Sparse Matched Filter Receiver operates over \(L\) pulses and involves calculating a vector \(\mathbf{h}(t;l)\) with complexity \(\mathcal{O}(K)\) per time instance \(t\), leading to \(\mathcal{O}(N_s K)\) total complexity. Each pulse involves \(Q\) sub-pulses. For each sub-pulse, inverting an \(M \times M\) matrix and multiplying it by a vector contributes \(\mathcal{O}(M^3 + M^2N_s)\). Processing each antenna element involves \(\mathcal{O}(N_s^2)\) operations, and other steps add \(\mathcal{O}(M^2 + N_s M)\). The overall complexity is dominated by \(\mathcal{O}(L Q M N_s^2) \approx \mathcal{O}(N_s^2)\), highlighting \(N_s^2\) as the most computationally demanding aspect.

The complexity of the modified Fuzzy C-Means (FCM) clustering algorithm includes three phases: clustering, equalizing cluster sizes, and adjusting labels.
\begin{itemize}
\item Phase I: Iterates until convergence, with each iteration involving \(\mathcal{O}(L \Phi_{T_l}^2)\) operations for membership calculations and center updates. Total complexity is \(\mathcal{O}(T L \Phi_{T_l}^2)\).
\item Phase II: Ensures clusters have equal sizes with a complexity of \(\mathcal{O}(L \Phi_{T_l}^2)\).
\item Phase III: Standardizes and adjusts cluster centers with complexity \(\mathcal{O}(\Phi_{T_l}^2)\) per iteration, totaling \(\mathcal{O}(T' \Phi_{T_l}^2)\).
\end{itemize}

Combining all phases, the overall complexity is \(\mathcal{O}(T L \Phi_{T_l}^2)\), reflecting linear dependence on data points \(L\) and quadratic dependence on cluster centers \(\Phi_{T_l}\), with iterative processes adding to the computational load.

%
% CONTINUE FROM HERE
%
\vspace{-1ex}
%============================================
\section{Performance Analysis And Discussion}
\label{Results}
To assess the performance of our method, we conducted simulations across diverse scenarios utilizing a \ac{MIMO} radar system within a wiretap channel setup. 
This involved simulating interactions between two legitimate partners, Alice and Bob, alongside an eavesdropper, identified as Eve.

Using the simulation results, we analyze key metrics including the secret bit disagreement rate, achievable bit rate, bit error rate (BER), and radar ambiguity function.
The simulation parameters are as follows: \(f_c = 10\) GHz, \(BW = 200\) MHz, \(f_s = 400\) MHz, \(K = 10\), \(\Delta_{f_L} = 50\) MHz, \(T_p = 10\) µs, \(\tau = 2\) µs, \(\Delta_t = 0.2\) µs, \(N = 8\), \(\Phi_{T_L}\) and \(\Phi_{f_L}\) take values in \(\{2, \underline{\mathbf{4}}, 8, 16\}\), \(\Delta_f = 5\) MHz, \(M \in \{1, 2, \dots, \underline{\mathbf{8}}, 9, 10\}\), \(J_{ASK} = 2\), and { \(J_{PSK} \in \{2, 4, 8\}\)}. The default values for these parameters are highlighted for easy reference.

\vspace{-2ex}
\subsection{Achievable Bit Rate}
Achievable bit rate refers to the maximum data transmission rate over a communication channel, expressed in bits per unit time, which is determined by PRF, $M$, $K$, $Q$, $J_{PSK}$ and $J_{ASK}$ for different embedding schemes as
\begin{equation}
\label{eq:Phase_Achievable_Bit_Rate}
R_{\text{ph}} = \text{PRF} \times Q \times \left ( M \log_2 J_{PSK} \right),  
\end{equation}
\begin{equation}
\label{eq:Amplitude_Achievable_Bit_Rate}
R_{\text{amp}} = \text{PRF} \times Q \times \left ( M \log_2 J_{ASK} \right).
\end{equation}

In turn, index modulation enhances the bit rate by selecting $M$ indices from a pool of $K$ indices as
\begin{equation}
\label{eq:Spatial_Index_Achievable_Bit_Rate}
R_{\text{sim}} = \text{PRF} \times Q \times  \left\lfloor \log_2 \left[\binom{K}{M} \times M! \right]\right\rfloor,   
\end{equation}
\begin{equation}
\label{eq:Hybrid_Achievable_Bit_Rate}
R_{\text{hyb}} = R_{\text{sim}} + \text{PRF} \times Q \times M  \left\lfloor\log_2 (J_{ASK} J_{PSK})\right\rfloor.  
\end{equation}

\begin{figure}[H]
\centering
\includegraphics[width=\columnwidth]{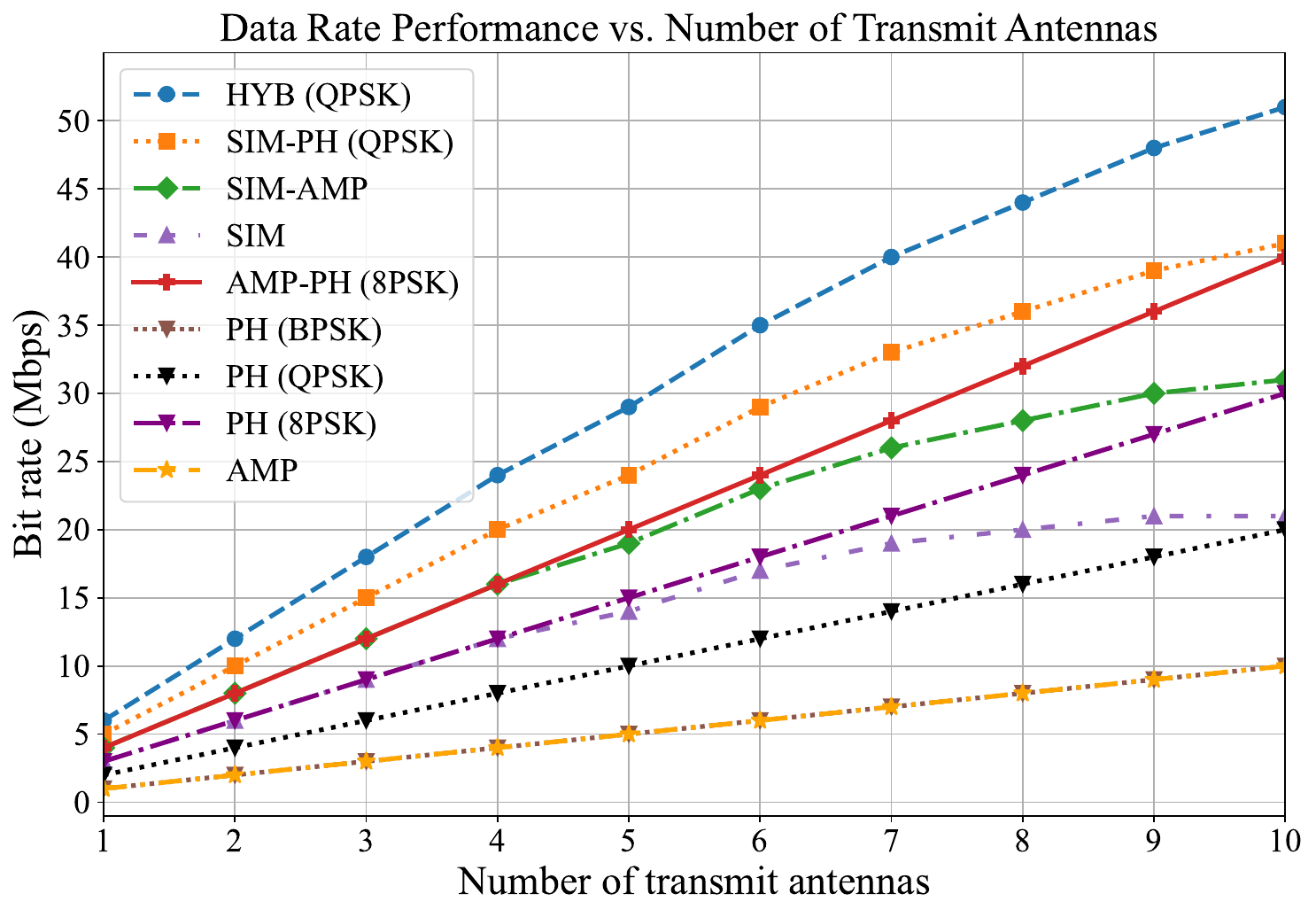}
\vspace{-5ex}
\caption{Achievable bit rates v.s the number of transmit antennas, $M$, for different information embedding schemes and { PSK constellation sizes.}}
\label{fig:Bit_Rate}
\end{figure}

Figure \ref{fig:Bit_Rate} shows the achievable bit rates for different information embedding schemes plotted against the number of transmit antennas, $M$. The bit rates for \ac{AMP} and \ac{PH} schemes increase linearly with $M$, following \eqref{eq:Phase_Achievable_Bit_Rate} and \eqref{eq:Amplitude_Achievable_Bit_Rate}. However, when \ac{AMP} uses a smaller constellation size ($J_{ASK}$) than \ac{PH}, resulting in a lower achievable rate.
It is seen that \ac{SIM} alone achieves a higher data rate compared to \ac{AMP} and\ac{PH} { (BPSK-QPSK)} schemes, with a logarithmic growth as described by \eqref{eq:Spatial_Index_Achievable_Bit_Rate}. The bit rate increases as $M$ grows from $1$ to $K/2$, but decreases from $K/2$ to $K$ due to the behavior of the term $\binom{K}{M}$.
For scenarios with limited \ac{TX} antennas, using $K/2$ antennas offers a good balance between achievable bit rate and resource utilization. Additionally, combining \ac{SIM} with \ac{AMP} or \ac{PH} schemes can further improve the rate.
In turn, the HYB method, which combines phase, amplitude, index, and spatial modulation, achieves the highest bit rate among all schemes, making it suitable for high data rates \ac{ISAC} applications.
\vspace{-2ex}

\vspace{-2ex}
\subsection{Bit Error Rate}
\begin{figure}[H]
\centering
\includegraphics[width=\columnwidth]{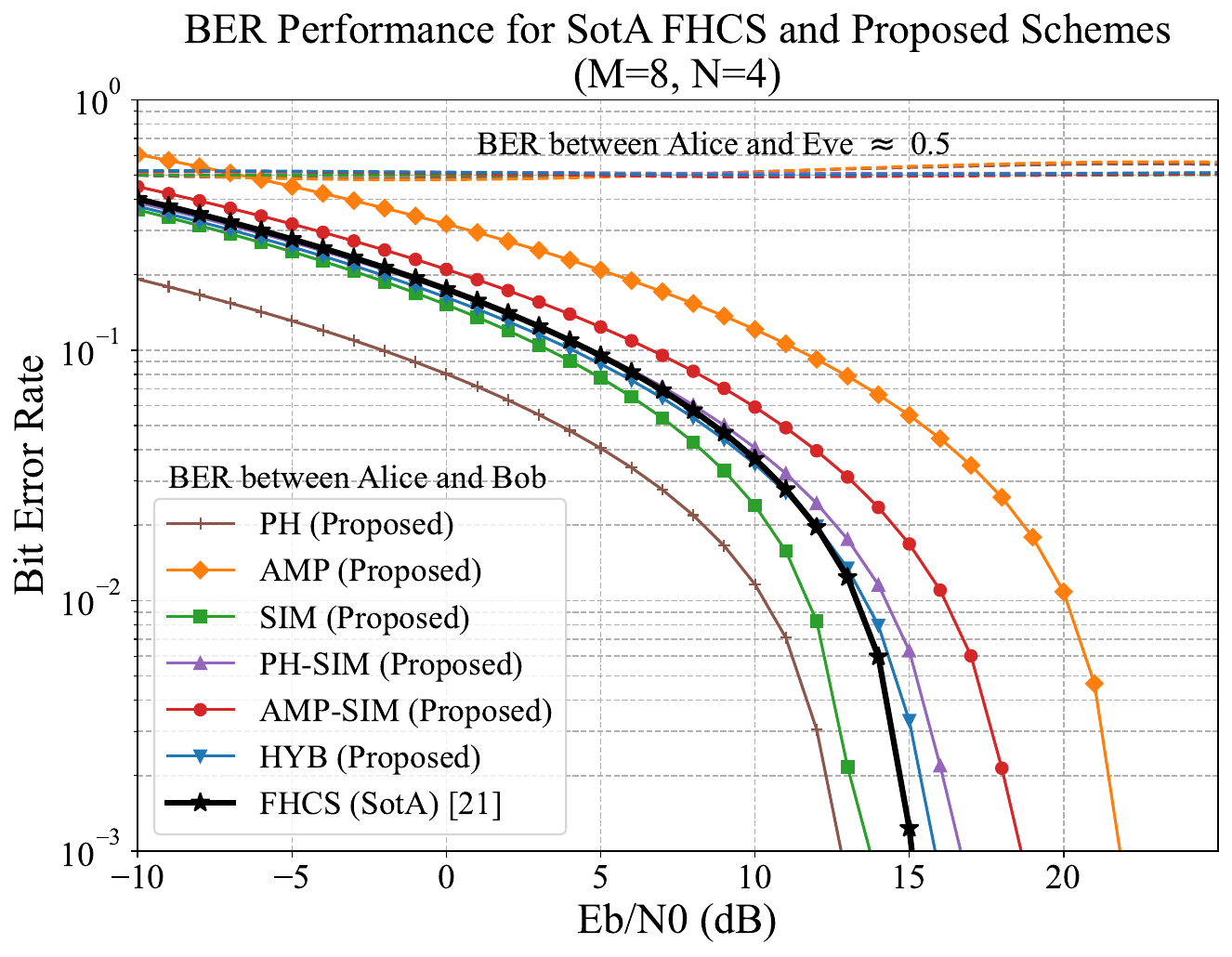}
\caption{\ac{BER} v.s { Eb/N0 (dB)} for communication between Alice and Bob (A \& B) and Alice and Eve (A \& E) across SotA FHCS and various individual information embedding schemes.}
\label{fig:Bit_Error_Rate}
\end{figure}

The \ac{BER} is a metric that quantifies the proportion of bits in the resulting key from Alice and Bob's protocol that do not align. This measure can also be assessed from Eve's perspective, ideally aiming for around $50\% $ to indicate optimal security \cite{AldaghriTIFS_2020}.
Fig. \ref{fig:Bit_Error_Rate} illustrates the \ac{BER} performance between Alice and Bob and also Alice and Eve for various communication schemes under different \ac{SNR}s.
Notably, for all methods, the \ac{BER} reaches a value close to 0.5 regardless of \ac{SNR}.
This is because Eve, lacking knowledge of the secret sequences $\mathbf{\Gamma}_T$ and $\mathbf{\Gamma}_f$, must guess them to eavesdrop.
This effectively limits the achievable information gain for eavesdroppers, enhancing data privacy. 
Additionally, all methods achieve near-perfect communication between Alice and Bob, with the \ac{BER} approaching zero for \ac{SNR}s up to around 18 dB.

This excellent performance is attributed to the use of matched filtering and \ac{OMP} techniques on the receiver side, both known for their high noise resistance.
The \ac{PH} scheme demonstrates superior performance due to its use of identical $\mathbf{\Gamma}_T$ and $\mathbf{\Gamma}_f$ sequences at both transmitter and receiver.
Furthermore, since no information is embedded in the phase or amplitude of the chips, \ac{SIM} also achieves good performance, which is because the employed 1-sparse \ac{OMP} receiver offers strong resistance to noise in the frequency domain.
However, { \ac{AMP}} is more susceptible to noise than other methods, because the changes in amplitude are more easily distorted during transmission compared to changes in phase.
Moreover, the \ac{BER} of \ac{PH}-\ac{SIM} and HYB schemes are comparable, which is influenced by the chosen values of parameters like $J_{PSK}$, $K$, $M$, and $Q$.
For instance, increasing $J_{PSK}$ might elevate the \ac{BER} of \ac{PH}-\ac{SIM}.
Notice that errors introduced by \ac{SIM} scheme also impact the overall \ac{BER} in both \ac{PH}-\ac{SIM} and \ac{AMP}-\ac{SIM} methods, which explains their weaker performance compared to other schemes.
It should also be noted, however, that the HYB method outperforms \ac{AMP}-\ac{SIM} because the incorporated \ac{PH} scheme helps reduce the total \ac{BER}.

{
The figure illustrates that our proposed sparse MF receiver effectively recovers the HYB scheme, achieving a comparable data rate and nearly identical \( E_b/N_0 \) performance to the \ac{SotA} \ac{FHCS} scheme. Notably, \ac{FHCS} relies on an exhaustive search through all possible combinations of \( M \) frequency indices selected from \( K \) available indices, in addition to \( M! \) permutations. This approach results in extremely high computational complexity, significantly limiting its practical applicability (\cite{BaxterTSP_2022} - issues and open problem). In contrast, our sparse MF receiver offers a more practical and computationally efficient alternative without compromising performance.

The secrecy rate in the wiretap channel model measures secure information transfer from Alice to Bob while limiting leakage to Eve. The secrecy capacity is considered as  
\begin{equation}
    C_s = \max_{P(\mathbf{X}^{(Alice)})} \left[ I(\mathbf{X}^{(\text{Alice})}; \mathbf{R}^{(\text{Bob})}) - I(\mathbf{X}^{(\text{Alice})}; \mathbf{R}^{(\text{Eve})}) \right]^+,
\end{equation}
where mutual information \( I \) has an inverse relationship with the \ac{BER}. Therefore, a lower \ac{BER} between Alice and Bob increases \( I(\mathbf{X}^{(\text{Alice})}; \mathbf{R}^{(\text{Bob})}) \), while a higher \ac{BER} for Eve (ideally 0.5) decreases \( I(\mathbf{X}^{(\text{Alice})}; \mathbf{R}^{(\text{Eve})}) \). This improves secrecy, ensuring secure communication.
}
\vspace{-3ex}
\subsection{Ambiguity Function}
\label{subsec:AF_Evaluation_Results}
The \ac{AF} is a powerful and efficient tool for analyzing and designing radar signals.
However, due to the random agility parameters, the \ac{AF} of \ac{RFPA} signals is randomly distributed on the delay-Doppler plane.
Consequently, it is essential to analyze the statistical characteristics of the \ac{AF} to gain insights into its behavior and performance.
The width of the main lobe and the height of the side lobes in the AF are crucial for radar signal analysis. 
%This analysis helps understand the variability and reliability of the radar signals, ultimately contributing to better \ac{ISAC} waveform design.
%
%Hence,let the \ac{AF} obtained in the \(i^{\text{th}}\) simulation be denoted as \(\chi_i(\tau, \nu)\). 
%
%The simulated expectation and variance after conducting a total of \(I\) Monte Carlo simulations are given by
%
%\begin{align}
%\label{eq:Expectation_variance_AF}
%&\mathbb{E}(\chi(\tau, \nu)) = \frac{1}{I}\sum\limits_{i=0}^{I-1}\chi_i(\tau, \nu),\\
%&\mathbb{V}ar (\chi(\tau, \nu)) = \frac{1}{I-1}\sum\limits_{i=0}^{I-1} |\chi_i(\tau, \nu) - \mathbb{E}(\chi(\tau, \nu))|^2.
%\end{align}

\begin{figure}[H]
\centering
\includegraphics[width=\columnwidth]{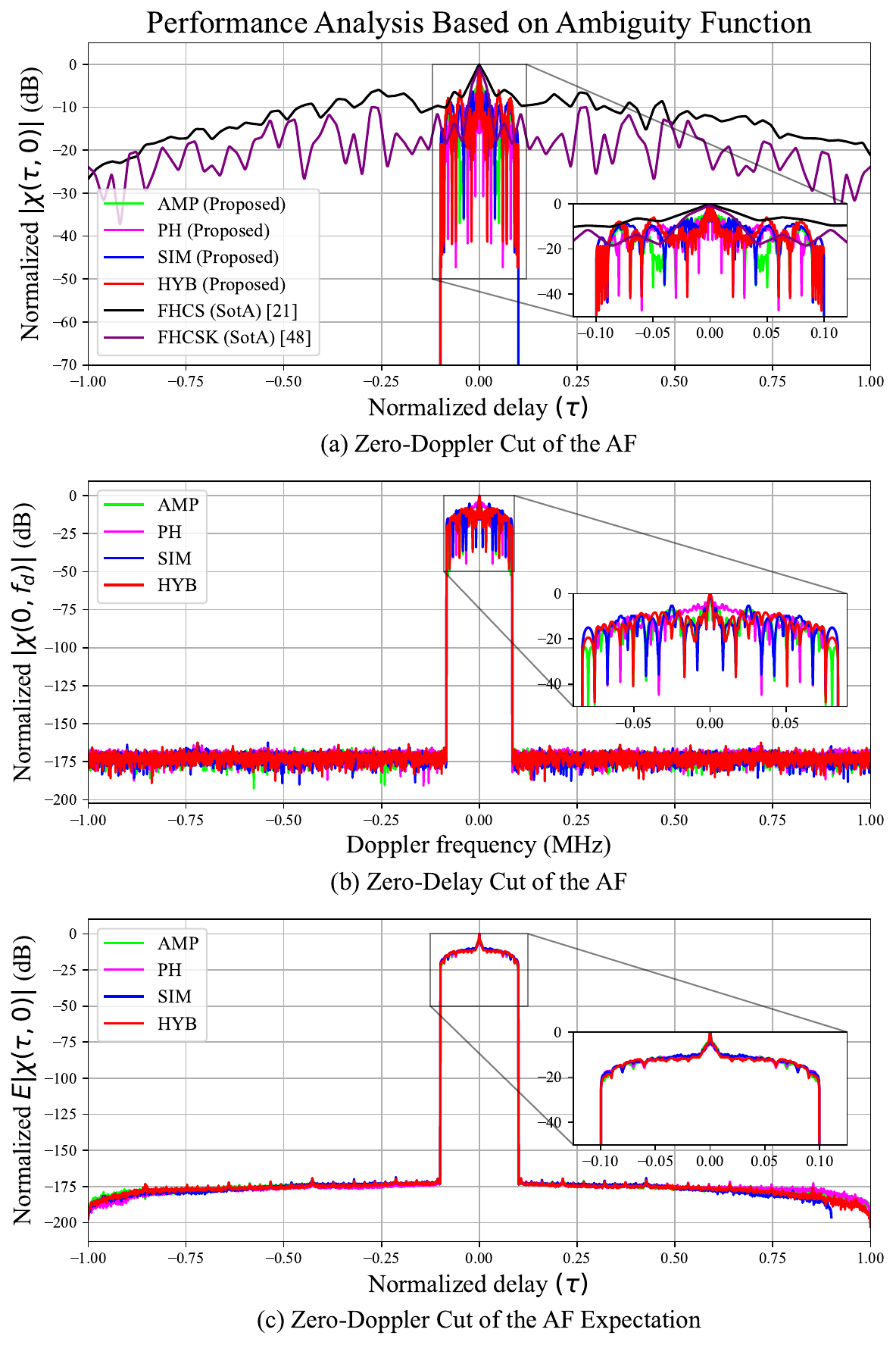}
\vspace{-5ex}
\caption{{ The \ac{AF} and its expectation for various information embedding schemes.  
(a) Zero-Doppler cut of the AF for SotA FHCS, FHCSK, and Proposed Schemes.  
(b) Zero-delay cut of the AF.  
(c) Zero-Doppler cut of the AF expectation.
}}
\label{fig:AF_plots}
\vspace{-3ex}
\end{figure}

{
Figure \ref{fig:AF_plots} illustrates the zero-Doppler and zero-delay cuts of the ambiguity function (AF), along with its expectation for various information embedding schemes.}
As shown in Figure \ref{fig:AF_plots} (A), the \ac{SotA} \acp{AF}, such as FHCS \cite{BaxterTSP_2022} and FHCSK \cite{Eedara_2022}, suffer from wide main lobes, high sidelobes, and ambiguities, negatively impacting target detection and clutter suppression.

{
 Embedding information in fast time amplifies these issues, as varying FH codes increase sidelobes and reduce suppression effectiveness. To overcome these challenges, we employ \ac{RFPA} schemes for \ac{ISAC} waveform design. Random PRI and frequency variables enhance range resolution, clutter resistance, and velocity estimation. 
Although the proposed HYB and \ac{SIM} schemes exhibit larger sidelobes compared to \ac{AMP} and \ac{PH}, their main lobe width and sidelobe heights are significantly smaller than the \ac{SotA} FHCS and FHCSK schemes.

Figure \ref{fig:AF_plots} (B) illustrates the zero-delay cut of the AF for our proposed \ac{RFPA} schemes, highlighting superior velocity estimation and resolution, along with better clutter suppression. Although the HYB and \ac{SIM} schemes show larger sidelobes due to varying FH codes, they still outperform the \ac{SotA} in accuracy. 
%\vspace{-0.5cm}
Figure \ref{fig:AF_plots} (C) displays the zero-Doppler cuts of the AF expectations for our proposed schemes, showing sharp main lobes and suppressed sidelobes, ensuring precise range estimation and robust performance in cluttered conditions.}

\subsection{Entropy}

{ In order to evaluate sensing secrecy, existing metrics in the literature are often context-specific and scenario-dependent. In contrast, since our focus is on generating strong secrets that remain unpredictable to passive adversaries, we employ metrics like Bit Disagreement Rate and Entropy, aligning with information-theoretic approaches for secret key generation.}

Entropy refers to the measure of unpredictability or randomness in the generated secret bits. 
{ In the \ac{RFPA} secret generation algorithm described in \ref{RFPA_Secret_Generation}, we employed the proposed vector quantization technique, which maximized achievable entropy compared to traditional scalar quantization \cite{Gyorgy_2000}. 
By generating secrets based on the \ac{CIR} shared between Alice and Bob, we used these as pseudo-random sequences in the \ac{RFA} and \ac{RPA} methods. Higher entropy is critical in this context as it directly enhances the unpredictability and randomness of the pseudo-random sequences, thereby strengthening both security and privacy \cite{Mukherjee_2014}.  
Increased entropy makes the sequences more resistant to attacks from eavesdroppers or passive adversary radars. This enhanced randomness improves \ac{AF} performance and radar estimation accuracy, which are crucial for the system's overall effectiveness.}

\begin{figure}[H]
\centering
\includegraphics[width=\columnwidth]{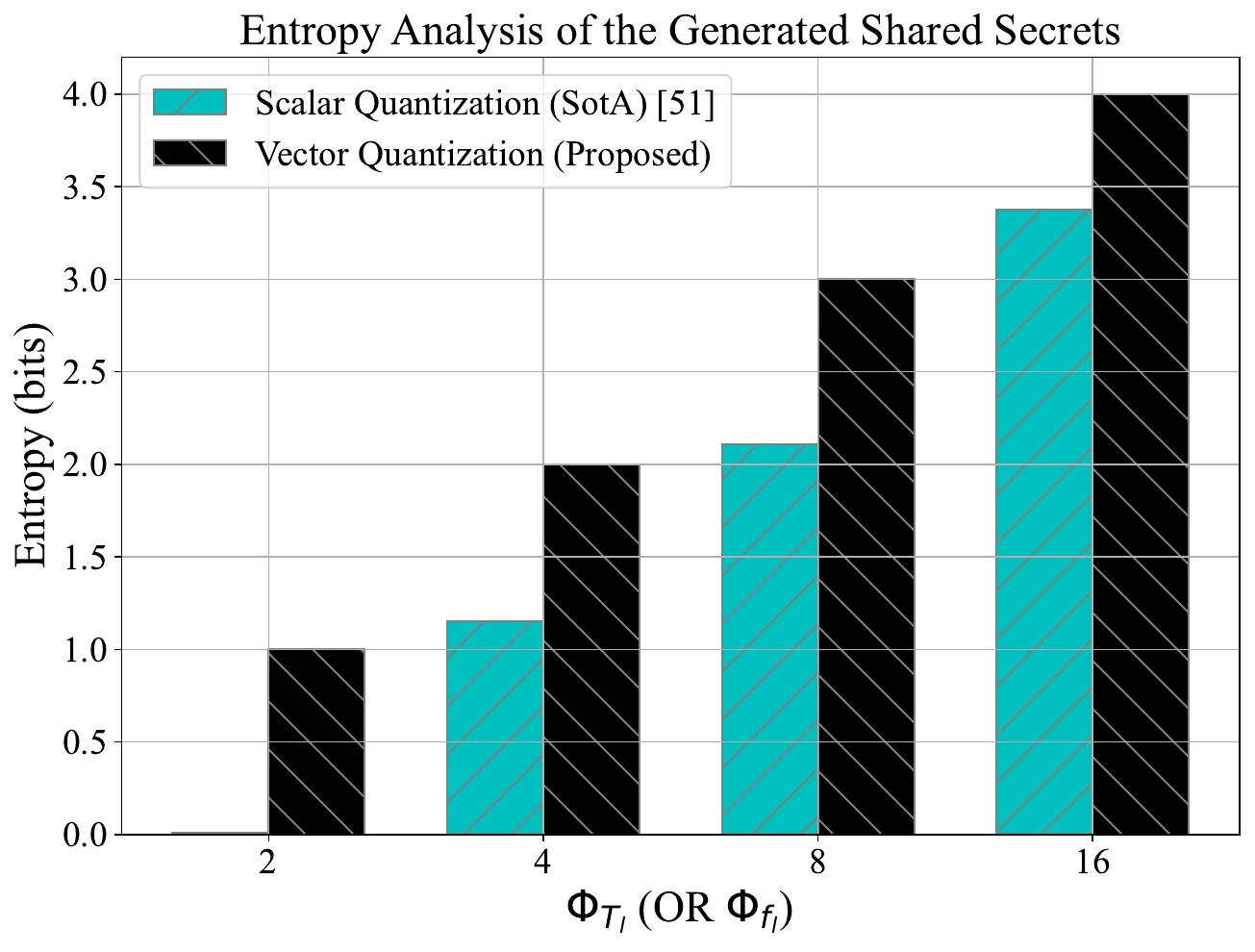}
\vspace{-0.8cm}
\caption{Entropy versus the number of quantization levels $\Phi_{T_l} = \Phi_{f_l} = \{2, 4, 8, 16\}$ for the proposed VQ and traditional \ac{SQ} techniques.}
\label{fig:Entropy}
\vspace{-3ex}
\end{figure}

{
Since the final shared secret \(\mathbf{\Gamma}_T\) obfuscates the PRI and \(\mathbf{\Gamma}_f\) obscures the Doppler frequency, they significantly complicate passive adversaries' ability to estimate the target's range and velocity. As a result, higher entropy lowers the probability of Eve successfully guessing these parameters.
}

In Fig. \ref{fig:Entropy}, we analyze the impact of our proposed scheme on entropy, comparing it with a traditional \ac{SQ} with quantization levels $\Phi_{T_l}$ or $\Phi_{f_l}$ ranging from 4 to 16. 
Our scheme consistently achieves the maximum entropy of \(\log_2 \Phi_{T_l}\) or \(\log_2 \Phi_{f_l}\), regardless of the SNR. 
This increased entropy is due to enhancements in the Fuzzy C-means algorithm, which ensures an equal distribution of members within each cluster, leading to equal probability for each quantization level. 
In contrast, traditional \ac{SQ}, which quantizes each channel sample independently, does not guarantee this equal distribution, resulting in significantly lower entropy.

\vspace{-3ex}
\subsection{Secret Bit Disagreement Rate}
Secret \ac{BDR}, serves as an indicator of the disparity between the secret data acquired independently by Alice and Bob {before} any reconciliation process. 
It stands as a pivotal metric evaluating the efficacy of reciprocity enhancement techniques and quantization algorithms. 
\vspace{-2ex}
\begin{figure}[H]
\centering
\includegraphics[width=\columnwidth, height=0.7\columnwidth]{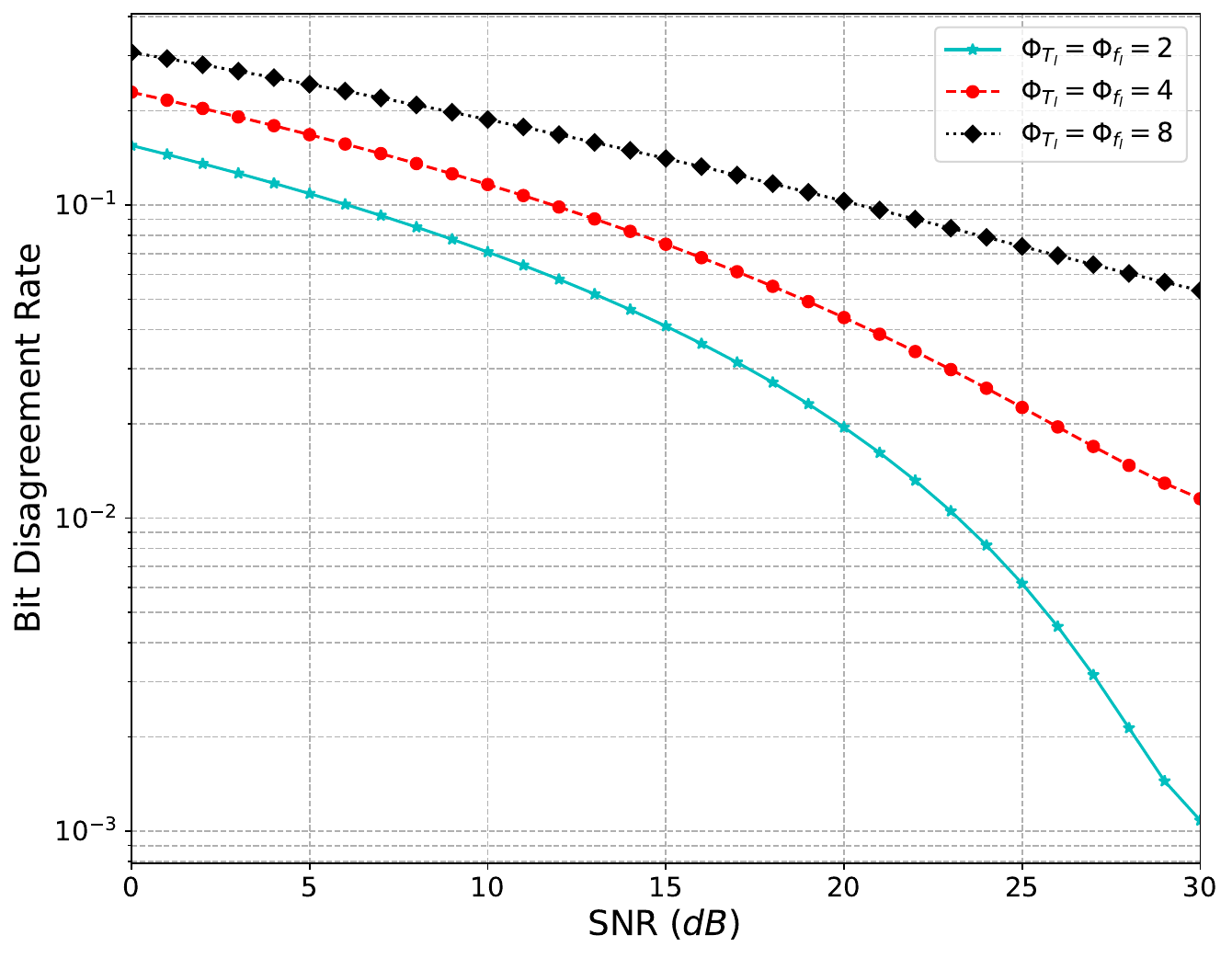}
\vspace{-0.8cm}
\caption{\ac{BDR} v.s \ac{SNR} (dB) for communication between Alice and Bob for various {numbers} of quantization levels $\Phi_{T_l} = \Phi_{f_l} = \{2, 4, 8\}$ for the proposed \ac{VQ} method.}
\label{fig:BDR}
\end{figure}

%
%This rate is formally defined as the ratio of disagreeing bits to the total number of bits, providing valuable insights into the robustness of the key generation process \cite{WangITJ_2024} as
%
%\begin{equation} 
%\label{eq:BDR}
%\begin{aligned}
%
%{\text{BDR}_T} &= \frac {\mathbf{\Gamma}_T^A \oplus \mathbf{\Gamma}_T^B}{\text {length}(\mathbf{\Gamma}_T)},\\
%
%{\text{BDR}_f} &= \frac {\mathbf{\Gamma}_f^A \oplus \mathbf{\Gamma}_f^B}{\text {length}(\mathbf{\Gamma}_f)},
%\end{aligned}
%\end{equation}
%
%where $(\mathbf{\Gamma}_T^A \oplus \mathbf{\Gamma}_T^B)$ and $(\mathbf{\Gamma}_f^A \oplus \mathbf{\Gamma}_f^B)$ represent the count of corresponding Gray bits in $\mathbf{\Gamma}_T$ and $\mathbf{\Gamma}_T$, respectively, where Alice and Bob disagree on, and $\text{length}(\mathbf{\Gamma}_T)$ and $\text{length}(\mathbf{\Gamma}_f)$ stand for the total bit length of the secrets.
%

Fig. \ref{fig:BDR} shows the \ac{BDR} versus \ac{SNR} in dB for communication between Alice and Bob. 
The plot compares \ac{BDR} for different numbers of quantization levels, $\Phi_{T_l} = \Phi_{f_l} = \{2, 4, 8\}$, for the proposed \ac{VQ} method.
As can be seen, the \ac{BDR} increases with the number of quantization levels.
This occurs because dividing the \ac{CIR} data points into more clusters increases the likelihood of points with slight noise being assigned to different clusters, which leads to a larger error between the secrets generated by Alice and Bob.
It is important to note that{,} unlike many other approaches, the initial proposed \ac{VQ} algorithm avoids communication overhead, which comes at the cost of a some \ac{BDR}.

\vspace{-2ex}
\section{Conclusion}
\label{Conclusion}

In this paper, we have addressed the critical need for enhancing the security and privacy of \ac{ISAC} systems. 
By introducing \ac{RFA} and \ac{RPA} techniques, we have developed a robust framework that secures data transmission and protects radar sensing information from unauthorized access. 
Our proposed \ac{RFPA} techniques effectively obscure Doppler frequency and pulse start times, complicating adversarial efforts to estimate target locations and velocities. 
The ambiguity function analysis confirms that our waveforms offer superior range and velocity resolution while minimizing clutter effects.
Additionally, we have presented a hybrid information embedding method combining \ac{ASK}, \ac{PSK}, \ac{IM}, and \ac{SM}, which significantly enhances the achievable bit rate, making our solution suitable for high-data-rate \ac{ISAC} applications.
The design of a sparse-matched filter receiver ensures efficient decoding with low computational complexity, maintaining a low \ac{BER} even in challenging conditions. 
Furthermore, the novel \ac{CRKG}-based \ac{RFPA} secret generation scheme enhances security by generating high-entropy, random codes without the need for a coordinating authority.
Our simulation results underscore the efficacy of our proposed methods, demonstrating notable improvements in communication performance, radar sensing accuracy, and system security. 
{
However, to enhance the security of \ac{ISAC} systems, it is crucial to investigate threats and attacks from both active and passive adversaries, especially those with advanced resources.  
Comprehensive analysis, advanced eavesdropping models, and machine learning-based real-time anomaly detection are necessary to strengthen robustness and mitigate risks at the physical layer.  
Additionally, \ac{FH} sequence optimization can significantly improve sensing performance and anti-jamming capabilities. Future research should explore whether this optimization can be applied collaboratively on both legitimate partners, Alice and Bob, in a way that remains uncorrelated with Eve’s optimizers, further enhancing security.  
Future work should also focus on safeguarding target privacy in \ac{ISAC} use cases, where telecommunication signals typically take precedence over radar signals.
}

\vspace{-1ex}
\section*{Acknowledgment}
This work was funded by the German Federal Ministry of Education and Research (grant 16KISK231 and grant 16KIS1399), the German Research Foundation (Germany's Excellence Strategy–EXC2050/1–ProjectID 390696704–Cluster of Excellence CeTI of Dresden, University of Technology), and {based on} the budget passed by the Saxon State Parliament.

%\ifCLASSOPTIONcaptionsoff
%\newpage
%\fi
%\bibliographystyle{IEEEtran}
%\bibliography{main.bib} 

\begin{thebibliography}{10}
\providecommand{\url}[1]{#1}
\csname url@samestyle\endcsname
\providecommand{\newblock}{\relax}
\providecommand{\bibinfo}[2]{#2}
\providecommand{\BIBentrySTDinterwordspacing}{\spaceskip=0pt\relax}
\providecommand{\BIBentryALTinterwordstretchfactor}{4}
\providecommand{\BIBentryALTinterwordspacing}{\spaceskip=\fontdimen2\font plus
\BIBentryALTinterwordstretchfactor\fontdimen3\font minus
  \fontdimen4\font\relax}
\providecommand{\BIBforeignlanguage}[2]{{%
\expandafter\ifx\csname l@#1\endcsname\relax
\typeout{** WARNING: IEEEtran.bst: No hyphenation pattern has been}%
\typeout{** loaded for the language `#1'. Using the pattern for}%
\typeout{** the default language instead.}%
\else
\language=\csname l@#1\endcsname
\fi
#2}}
\providecommand{\BIBdecl}{\relax}
\BIBdecl


\bibitem{Liu_JSC22}
F.~Liu \emph{et al.}, ``{I}ntegrated {S}ensing and {C}ommunications: {T}oward {D}ual-{F}unctional {W}ireless {N}etworks for {6G} and {B}eyond,'' \emph{IEEE Journal on Selected Areas in Communications}, vol.~40, no.~6, 2022.

\bibitem{WangITJ2022}
J.~Wang \emph{et al.}, ``{I}ntegrated {S}ensing and {C}ommunication: {E}nabling {T}echniques, {A}pplications, {T}ools and {D}ata {S}ets, {S}tandardization, and {F}uture {D}irections,'' \emph{IEEE Internet of Things Journal}, no.~23, 2022.

\bibitem{Wei_ITJ23}
Z.~Wei \emph{et al.}, ``{I}ntegrated {S}ensing and {C}ommunication {S}ignals {T}oward {5G-A} and {6G}: {A} {S}urvey,'' \emph{IEEE Internet of Things Journal}, vol.~10, no.~13, 2023.

\bibitem{RanasingheTWC2024}
K.~R.~R. Ranasinghe \emph{et al.}, ``{J}oint {C}hannel, {D}ata and {R}adar {P}arameter {E}stimation for {AFDM} {S}ystems in {D}oubly-{D}ispersive {C}hannels,'' \emph{IEEE Transactions on Wireless Communications}, pp. 1--1, 2024.

\bibitem{ISACMarket2023}
``{G}lobal {I}ntegrated {S}ensing and {C}ommunication ({ISAC}) {M}arket by {T}ype ({S}emi-{ISAC}, {UAV}-enabled {ISAC}), by {A}pplication ({C}ar, {D}rone), by {G}eographic {S}cope and {F}orecast,'' Verified Market Reports, 2023.

\bibitem{KaiqianISAC2023}
K.~Qu \emph{et al.}, ``{P}rivacy and {S}ecurity in {U}biquitous {I}ntegrated {S}ensing and {C}ommunication: {T}hreats, {C}hallenges and {F}uture {D}irections,'' 2023.

\bibitem{LiComMag2023}
X.~Li \emph{et al.}, ``{I}ntegrated {H}uman {A}ctivity {S}ensing and {C}ommunications,'' \emph{IEEE Communications Magazine}, vol.~61, no.~5, 2023.

\bibitem{ZhangNetwork2024}
Y.~Zhang \emph{et al.}, ``{AI} {E}mpowered {C}hannel {S}emantic {A}cquisition for {6G} {I}ntegrated {S}ensing and {C}ommunication {N}etworks,'' \emph{IEEE Network}, 2024.

\bibitem{ChenWCom2023}
M.~Chen \emph{et al.}, ``{G}uest {E}ditorial: {AI}-driven {T}heory, {T}echnology and {A}pplication for {S}ensing, {I}nteraction, and {D}igitalization in the {6G} {E}ra,'' \emph{IEEE Wireless Communications}, vol.~30, no.~3, 2023.

\bibitem{SuTWC_2021}
N.~Su, F.~Liu, and C.~Masouros, ``{S}ecure {R}adar-{C}ommunication {S}ystems with {M}alicious {T}argets: {I}ntegrating {R}adar, {C}ommunications and {J}amming {F}unctionalities,'' \emph{IEEE Transactions on Wireless Communications}, vol.~20, no.~1, 2021.

\bibitem{YuSecV2XISAC2023}
K.~Yu \emph{et al.}, ``{S}ecure {V2X} {C}ommunication: {A}n {I}ntegrated {S}ensing and {C}ommunication {P}erspective,'' 2023.

\bibitem{GunluJSAIT2023}
O.~G{\:u}nl{\:u}, M.~R. Bloch, R.~F. Schaefer, and A.~Yener, ``{S}ecure {I}ntegrated {S}ensing and {C}ommunication,'' \emph{IEEE Journal on Selected Areas in Information Theory}, vol.~4, 2023.

\bibitem{ylianttila6GSec2020}
M.~Ylianttila \emph{et al.}, ``6{G} {W}hite {P}aper: {R}esearch {C}hallenges for {T}rust, {S}ecurity and {P}rivacy,'' 2020.

\bibitem{MucchiOJCS2021}
L.~Mucchi \emph{et al.}, ``{P}hysical-layer {S}ecurity in {6G} {N}etworks,'' \emph{IEEE Open Journal of the Communications Society}, vol.~2, 2021.

\bibitem{LiuITJ2021}
J.~Liu \emph{et al.}, ``{P}ost-{Q}uantum {S}ecure {R}ing {S}ignatures for {S}ecurity and {P}rivacy in the {C}ybertwin-driven {6G},'' \emph{IEEE Internet of Things Journal}, vol.~8, no.~22, 2021.

\bibitem{KatsukiTIFS2023}
Y.~Katsuki, G.~T. F.~d. Abreu, K.~Ishibashi, and N.~Ishikawa, ``{N}oncoherent {M}assive {MIMO} with {E}mbedded {O}ne-{W}ay {F}unction {P}hysical {L}ayer {S}ecurity,'' \emph{IEEE Trans. Inf. Forensics Secur.}, vol.~18, 2023.

\bibitem{WuTWC2022}
K.~Wu \emph{et al.}, ``{I}ntegrating {S}ecure {C}ommunications into {F}requency {H}opping {MIMO} {R}adar with {I}mproved {D}ata {R}ate,'' \emph{IEEE Transactions on Wireless Communications}, vol.~21, no.~7, 2022.

\bibitem{RanasingheICASSP2024}
K.~R.~R. Ranasinghe, H.~S. Rou, and G.~T.~F. de~Abreu, ``{F}ast and {E}fficient {S}equential {R}adar {P}arameter {E}stimation in {MIMO}-{OTFS} {S}ystems,'' in \emph{ICASSP 2024 - 2024 IEEE International Conference on Acoustics, Speech and Signal Processing (ICASSP)}, 2024.

\bibitem{RanasingheWCNC2025}
K.~R.~R. Ranasinghe \emph{et al.}, ``{B}lind {B}istatic {R}adar {P}arameter {E}stimation for {AFDM} {S}ystems in {D}oubly-{D}ispersive {C}hannels,'' to appear in \textit{Proc. IEEE Wireless Commun. and Networking Conference (WCNC)}, 2025.

\bibitem{HoangWCL_2022}
L.~M. Hoang \emph{et al.}, ``{F}requency {H}opping {J}oint {R}adar-{C}ommunications with {H}ybrid {S}ub-pulse {F}requency and {D}uration {M}odulation,'' \emph{IEEE Wireless Communications Letters}, vol.~11, no.~11, 2022.

\bibitem{BaxterTSP_2022}
W.~Baxter \emph{et al.}, ``{J}oint {R}adar and {C}ommunications for {F}requency-hopped {MIMO} {S}ystems,'' \emph{IEEE Trans. Sig. Proc.}, vol.~70, 2022.

{\bibitem{HassanienRC_2017}
A. Hassanien \emph{et al.}, ``A {D}ual-{F}unction {MIMO} {R}adar-{C}ommunications {S}ystem {U}sing {F}requency-{H}opping {W}aveforms,'' in \emph{IEEE} {R}adar {C}onference ({RadarConf}), 2017.}

\bibitem{ChenTSP_2008}
C.-Y. Chen and P.~P. Vaidyanathan, ``{MIMO} {R}adar {A}mbiguity {P}roperties and {O}ptimization {U}sing {F}requency-{H}opping {W}aveforms,'' \emph{IEEE Transactions on Signal Processing}, vol.~56, no.~12, 2008.

\bibitem{LongTAES_2021}
X.~Long \emph{et al.}, ``{A}mbiguity {F}unction {A}nalysis of {R}andom {F}requency and {PRI} {A}gile {S}ignals,'' \emph{IEEE Transactions on Aerospace and Electronic Systems}, vol.~57, no.~1, 2021.

\bibitem{AngelosanteTSP_2010}
D.~Angelosante \emph{et al.}, ``{E}stimating {M}ultiple {F}requency-hopping {S}ignal {P}arameters via {S}parse {L}inear {R}egression,'' \emph{IEEE Transactions on Signal Processing}, vol.~58, no.~10, 2010.

\bibitem{LiuSPM_2023}
F.~Liu \emph{et al.}, ``{S}eventy {Y}ears of {R}adar and {C}ommunications: {T}he {R}oad from {S}eparation to {I}ntegration,'' \emph{IEEE Signal Processing Magazine}, vol.~40, no.~5, 2023.

{
\bibitem{DongyangVTC2017}
Xu, Dongyang \emph{et al.}, ``{A} {C}omparison of {N}orm {B}ased {A}ntenna {S}election and {R}andom {A}ntenna {S}election with {R}egard to {E}nergy {E}fficiency in {W}ireless {S}ystem with {L}arge {N}umber of {U}sers,'' in \emph{IEEE 86th Vehicular Technology Conference (VTC-Fall)}, 2017.

\bibitem{NiuTC2022}
Niu, Hong and Xiao, Yue and Lei, Xia and Xiao, Ming, ``{A} {C}omparison of {A}rtificial {N}oise {E}limination: {F}rom the {P}erspective of {E}avesdroppers,'' \emph{IEEE Transactions on Communications}, vol.~70, no.~7, 2022.

\bibitem{YehTWC2021}
Yeh, Chia-Yi and Knightly, Edward W., ``{E}avesdropping in {M}assive {MIMO}: {N}ew {V}ulnerabilities and {C}ountermeasures,'' \emph{IEEE Transactions on Wireless Communications}, vol.~20, no.~10, 2021.

\bibitem{Taneja2020}
{A. Taneja and N. Saluja}, ``{A} {C}omparison of {N}orm Based {A}ntenna {S}election and {R}andom {A}ntenna {S}election with {R}egard to {E}nergy {E}fficiency in {W}ireless {S}ystem with {L}arge {N}umber of {U}sers,'' \emph{SWCC}, vol.~10, no.~2, 2020.

\bibitem{Sanayei_2004}
Sanayei, S. and Nosratinia, A., ``{A}ntenna {S}election in {MIMO} {S}ystems,'' \emph{IEEE Communications Magazine}, vol.~42, no.~10, 2004.
}
\bibitem{EusticeWMCS_2015}
D.~Eustice, C.~Baylis, and R.~J. Marks, ``{W}oodward's {A}mbiguity {F}unction: {F}rom {F}oundations to {A}pplications,'' in \emph{Texas {S}ymposium on {W}ireless and {M}icrowave {C}ircuits and {S}ystems ({WMCS})}, 2015.

\bibitem{AlaeeICASSP_2019}
M.~Alaee-Kerahroodi \emph{et al.}, ``{D}esigning ({I}n)finite-{A}lphabet {S}equences {V}ia {S}haping the {R}adar {A}mbiguity {F}unction,'' in \emph{IEEE {I}nternational {C}onference on {A}coustics, {S}peech and {S}ignal {P}rocessing}, 2019.

\bibitem{ZhangISJ_2016}
J.~Zhang \emph{et al.}, ``{S}haping {R}adar {A}mbiguity {F}unction by $l$-{P}hase {U}nimodular {S}equence,'' \emph{IEEE {S}ensors {J}ournal}, vol.~16, no.~14, 2016.

\bibitem{SedivyCOMITE_2013}
P.~Sedivy, ``{R}adar {PRF} {S}taggering and {A}gility {C}ontrol {M}aximizing {O}verall {B}lind {S}peed,'' in \emph{Conference on {M}icrowave {T}echniques}, 2013.

\bibitem{LiuTSP_2014}
Y.~Liu \emph{et al.}, ``{F}undamental {L}imits of {HRR} {P}rofiling and {V}elocity {C}ompensation for {S}tepped-{F}requency {W}aveforms,'' \emph{IEEE {T}ransactions on {S}ignal {P}rocessing}, vol.~62, no.~17, 2014.

\bibitem{XudongICR_2011}
X.~Cao, H.~Fan, and X.~Wu, ``{E}CCM {P}erformance {A}nalysis of {I}nter-{P}ulses {F}requency {A}gility {A}pplication,'' in \emph{Proceedings of 2011 {IEEE} CIE {I}nternational {C}onference on {R}adar}, vol.~1, 2011.

\bibitem{BasarCM_2016}
E.~Basar, ``{I}ndex {M}odulation {T}echniques for {5G} {W}ireless {N}etworks,'' \emph{IEEE {C}ommunications {M}agazine}, vol.~54, no.~7, 2016.

\bibitem{RouTWC_2022}
H.~S. Rou, G.~T.~F. de~Abreu, H.~Iimori, D.~G. G., and O.~Gonsa, ``{S}calable {Q}uadrature {S}patial {M}odulation,'' \emph{IEEE {T}ransactions on {W}ireless {C}ommunications}, vol.~21, no.~11, 2022.

\bibitem{DavenportTIT_2010}
M.~A. Davenport and M.~B. Wakin, ``{A}nalysis of {O}rthogonal {M}atching {P}ursuit {U}sing the {R}estricted {I}sometry {P}roperty,'' \emph{IEEE {T}ransactions on {I}nformation {T}heory}, vol.~56, no.~9, 2010.

\bibitem{CaiTIT_2011}
T.~T. Cai and L.~Wang, ``{O}rthogonal {M}atching {P}ursuit for {S}parse {S}ignal {R}ecovery with {N}oise,'' \emph{IEEE {T}ransactions on {I}nformation {T}heory}, vol.~57, no.~7, 2011.

\bibitem{BagheriGIIS_2024}
G.~Bagheri, A.~K. Boroujeni, and S.~Köpsell, ``{M}achine {L}earning-based {V}ector {Q}uantization for {S}ecret {K}ey {G}eneration in {P}hysical {L}ayer {S}ecurity,'' in \emph{2024 {G}lobal {I}nformation {I}nfrastructure and {N}etworking {S}ymposium ({GIIS})}, 2024.

\bibitem{HanIS_2020}
Q.~Han \emph{et al.}, ``{V}ector {P}artitioning {Q}uantization {U}tilizing {K}-means {C}lustering for {P}hysical {L}ayer {S}ecret {K}ey {G}eneration,'' \emph{Information {S}ciences}, vol. 512, 2020.

\bibitem{HongTIFS_2017}
Y.-W.~P. Hong \emph{et al.}, ``{V}ector {Q}uantization and {C}lustered {K}ey {M}apping for {C}hannel-based {S}ecret {K}ey {G}eneration,'' \emph{IEEE {T}ransactions on {I}nformation {F}orensics and {S}ecurity}, vol.~12, no.~5, 2017.

\bibitem{ChenTWC_2022}
C.~Chen, ``{S}ample-grouping-based {V}ector {Q}uantization for {S}ecret {K}ey {E}xtraction from {A}tmospheric {O}ptical {W}ireless {C}hannels,'' \emph{IEEE {T}ransactions on {W}ireless {C}ommunications}, vol.~21, no.~11, 2022.

\bibitem{PengICACES_2022}
J.~Peng \emph{et al.}, ``{S}ecret {K}ey {G}eneration {U}sing {P}olar {C}ode-based {R}econcilation {M}ethod in 5{G},'' in \emph{International {C}onference on {A}dvanced {C}omputing and {E}ndogenous {S}ecurity}, 2022.

\bibitem{TalTIT_2015}
I.~Tal and A.~Vardy, ``{L}ist {D}ecoding of {P}olar {C}odes,'' \emph{IEEE {T}ransactions on {I}nformation {T}heory}, vol.~61, no.~5, 2015.

\bibitem{ShakibaSSP_2021}
M.~Shakiba-Herfeh and A.~Chorti, ``{C}omparison of {S}hort {B}locklength {S}lepian-{W}olf {C}oding for {K}ey {R}econcilation,'' in \emph{IEEE {S}tatistical {S}ignal {P}rocessing {W}orkshop ({SSP})}, 2021.

\bibitem{AldaghriTIFS_2020}
N.~Aldaghri and H.~Mahdavifar, ``{P}hysical {L}ayer {S}ecret {K}ey {G}eneration in {S}tatic {E}nvironments,'' \emph{IEEE {T}ransactions on {I}nformation {F}orensics and {S}ecurity}, vol.~15, 2020.
{
\bibitem{Eedara_2022}
Eedara, Indu Priya \emph{et al.}, ``{D}ual-{F}unction {F}requency-{H}opping {MIMO} {R}adar {S}ystem {W}ith {CSK} {S}ignaling,'' \emph{IEEE {T}ransactions on {A}erospace and {E}lectronic {S}ystems}, vol.~58, no.~3, 2022.

\bibitem{Gyorgy_2000}
Gyorgy, A. and Linder, T., ``{O}ptimal {E}ntropy-{C}onstrained {S}calar {Q}uantization of a {U}niform {S}ource,'' \emph{IEEE {T}ransactions on {I}nformation {T}heory}, vol.~46, no.~7, 2000.}

\bibitem{Mukherjee_2014}
Mukherjee, Amitav \emph{et al.}, ``{P}rinciples of {P}hysical {L}ayer {S}ecurity in {M}ultiuser {W}ireless {N}etworks: A {S}urvey,'' \emph{IEEE {C}ommunications {S}urveys and {T}utorials}, vol.~16, no.~3, 2014.



\end{thebibliography}


\end{document}